\documentstyle[eqsecnum,aps]{revtex}
\begin{document}

\draft

 \title{Deep inelastic scattering on the deuteron
in the Bethe-Salpeter formalism II:\\
 Realistic $NN$-interaction.}

\author{A.Yu.  Umnikov\thanks{INFN
Postdoctoral Fellow}\thanks{{\em On
leave from } INFN, Sezione di Perugia  and
Department of Physics, University of Perugia,
via A. Pascoli, Perugia, I-06100, Italy.}
and  F.C Khanna }

\address   {
    University of Alberta, Edmonton,
 Alberta T6G 2J1, Canada
and TRIUMF, 4004 Wesbrook Mall,
Vancouver, BC, Canada, V6T 2A3.}

\author{ L.P. Kaptari }

\address{
   Bogoliubov Laboratory  of Theoretical  Physics,
Joint Institute for Nuclear Research,
   Dubna, Russia.}

\date{\today}

\maketitle


\begin{abstract} We present a systematic study of  the leading
twist structure functions of the deuteron, $F_{2}^D$, $b_{1,2}^D$
and $g_1^D$ in a fully relativistic approach. Our study is based
on a realistic Bethe-Salpeter amplitude for the deuteron, which
is obtained as a solution to the homogeneous Bethe-Salpeter
equation with a realistic $NN$ kernel. Particular effort is made
to connect  the structure functions to the densities of the
appropriate charges and currents. This  allows for a systematic
comparison between the relativistic  and nonrelativistic
calculations, by  analysing the same densities in both
approaches. Thus, the sources of the relativistic effects in the
structure functions are understood  and clearly  distinguished
from variations  caused by the differences in the model
parameters. We present both the formalism and extensive
numerical calculations for all steps of our analysis.

We find that the nonrelativistic and relativistic calculations
are  qualitatively very much alike. However, three main features
systematically distinguish  a consistent relativistic approach
from the nonrelativistic one: (i) the binding effects are larger,
(ii) the effect of Fermi motion at high $x$ is stronger and (iii)
the  relativistic description of the structure functions
$b_{1,2}^D$  is fully consistent, unlike the nonrelativistic
approach, which is internally  inconsistent and violates the
fundamental sum rules.

\end{abstract}

\section{INTRODUCTION} \label{sec:introduction}

This paper is the second of the two papers devoted to a study of
the deep inelastic lepton  scattering (inclusive
electroproduction) on the deuteron within the Bethe-Salpeter (BS)
formalism.  The first paper~\cite{1}, presented a formal
approach  to the deep inelastic scattering within the BS
formalism. The emphasize was   on the self-consistency of the
method,  and the development of all aspects of the formalism so
that it could be applied to a study of the realistic  cases of
reactions.  Basic analysis was performed utilizing the  operator
product expansion method in the leading twist approximation.
Subsequently, the results were  obtained in the space of the
moments of the structure functions. Formal consistency of the
approach  was established and important sum rules were  proven.
However, all   numerical estimates were performed in the
``scalar deuteron'' model which is rather far from reality. At
the same time, it was stated that   the following tasks have to
be done in order to complete  the study. First,  explicit
formulas for the deuteron spin-averaged and spin-dependent
structure functions have to be derived, and, second,    the
structure functions have to be computed in  a realistic
meson-nucleon theory.  The present paper deals with both of the
tasks.

Utilizing the  realistic BS amplitude of the deuteron, we
calculate all the leading twist structure functions of the
deuteron, and analyze the manifestations of the relativistic
effects in these  structure functions.   Some of the preliminary
results, including solution of the  BS equation  with realistic
$NN$-interactions, were discussed
previously~\cite{urel,urel1,urel2}. However, this paper is not
just a simple sum of results reported elsewhere, nor is it
based solely on the direct application  of the formalism
developed  in ref.~\cite{1},   but rather presents a better
understanding of the formal and physical aspects of the process.

It is imperative to mention now  that  in recent years   a number
of studies of the deep inelastic scattering of leptons off the
deuteron have been done,  referring to various manifestations of
the relativistic features of the
deuteron~\cite{moff,mf2,mg1,kulg1,tokarev,lev}. These studies
certainly influenced our analysis  during extensive studies in
the time following the first publication~\cite{1}. Next, we found
it convenient to use the formalism developed in
refs.~\cite{fs,land,km} in the late 1970-th (see
also~\cite{msm}).   This formalism is  more convenient if we deal
with the  structure functions, rather than the moments of
structure functions (although it is mathematically equivalent, of
course). Indeed,    the formalism of ref.~\cite{1} allows us to
calculate the physical moments of the structure functions  using
the deuteron amplitude under the Wick rotation.  However, the
{\em numerical transform}  from the moments to the  structure
functions (the  inverse Mellin transform)  is  not a well-defined
operation. The new formalism allows for  numerical treatment of
the ``inverse Wick rotation'' of the Bethe-Salpeter amplitude of
the deuteron for an application to the calculation  of the
structure functions.

Another new distinguishing  feature of the present paper is the
special  accent on the systematic comparison of the relativistic
and non-relativistic  approaches. To do this,   we consider a
connection of the structure functions to the densities of
appropriate charges and currents, and then analyze these
densities  in both relativistic and nonrelativistic approaches.
In this way, we are able to trace the origins of the relativistic
effects and   demonstrate inconsistencies in the  nonrelativistic
approximations.    The paper contains a number of illustrations
for every step of  our analysis.

Finally, results for the structure functions published
previously  have been recalculated using the new method and some
of  them have been corrected slightly.

The paper is organized as follows. In Section~\ref{sec:anatomy},
we define the densities of the currents which are used  later in
the analysis of the deep inelastic scattering. For this purpose,
the realistic BS amplitude  for the deuteron is presented, and
the relativistic and nonrelativistic expressions for  densities
are defined and calculated within the realistic models. In
Section~\ref{sec:th} the relativistic  formalism of the approach
to the deep inelastic scattering on the deuteron is developed,
including    explicit  relativistic and nonrelativistic  formulae
for the  structure functions, and the sum rules for the structure
functions are analyzed.  Section~\ref{sec:num} contains results
of the calculations of the structure functions $F_2^D$,
$b_{1,2}^D$ and $g_1^D$ in the realistic models, both
relativistic and nonrelativistic. In Section~\ref{sec:sum} the
summary of results of the paper is presented.  Two Appendices
contain important  technical details.

\section{Anatomy of the Deuteron} \label{sec:anatomy}

\subsection{Realistic $NN$-interaction and Bethe-Salpeter
amplitude}

The basis of our approach to the relativistic description of the
deep inelastic  scattering off the deuteron is a  nucleon-nucleon
Bethe-Salpeter amplitude. General self-consistency of the
approach has been analysed in our  previous paper~\cite{1}. The
way to construct relevant matrix elements has been shown and
important sum rules have been proved, both  without any reference
to the particular model for the BS amplitude. A naive numerical
estimate has been done within the {\em   scalar deuteron} model
which is found to be in a qualitative agreement with the
nonrelativistic theory of the deep inelastic
scattering~\cite{akv,bls,unon}.

Apparently, for the realistic calculations of the deuteron
structure functions we need the {\em realistic} BS amplitude. The
realistic  BS amplitude of the deuteron is such an amplitude
which  provides us with a good description of bulk of the
deuteron properties. This is the same ideology as it is in
constructing the realistic wave function of the
deuteron~\cite{paris,bonn,gross2,tjond,gross}. The most
consistent way to obtain the realistic BS amplitude is to solve
dynamical problem within the {\em realistic model} of the
$NN$-interaction. The realistic $NN$-interaction still cannot be
derived from the underlying fundamental theory, QCD, or its
nonrelativistic approximations,  such as the chiral perturbation
theory. Although some progress has been achieved in  recent
years~\cite{nn1,nn2}. Alternatively, the parameters of the
$NN$-interaction can be fixed  from the available experimental
data on nucleon-nucleon scattering (phase shifts analysis), if
the general form of interaction is  somehow fixed. This is a
traditional way of  action~\cite{paris,bonn,tjonnn,gross}.
Unfortunately, the parametrization of the $NN$-interaction is
dependent on the choice of a dynamical equation. In spite of
certain similarity of the sets of parameters in all realistic
models there is no universal potential for all nonrelativistic
and relativistic potentials.

Only one parametrization of the $NN$-interaction  is available
for the BS equation, the parametrization of Fleischer and
Tjon~\cite{tjonnn}. This parametrization   is probably already
due for a revision~\cite{grossams}, incorporating  new data on
the nucleon-nucleon phase shift analysis~\cite{nnexp}. However,
this parametrization has been used as the basis for  our recent
calculation of the BS amplitude of the deuteron~\cite{urel}. The
meson parameters (masses, coupling constants, cut-off
parameters)  have been taken to be the same as  in
ref.~\cite{tjonnn,tjond}, except for the coupling constant of the
scalar $\sigma$-meson, which has been adjusted  to provide a
numerical solution of the  homogeneous BS equation. The chosen
set of   meson parameters   requires some minor adjustment, in
view of two circumstances. First, here we use a simplified form
of the propagator of the vector mesons, omitting
$k_{\mu}k_{\nu}/\mu_B^2$-term. Second, the different   numerical
procedures in solving the eigenvalue problem for the BS equation
can also  affect the value of parameters.

A detailed formalism preceding the numerical solution of the BS
equation for the deuteron is presented in ref.~\cite{1}. The
Fredholm system of "Wick rotated" equations with all meson
exchanges is solved by an iteration procedure with two
dimensional gaussian integration~(see e.g. ref.~\cite{ukspe}).
Eight components of the BS amplitude of the deuteron:
\begin{eqnarray} \psi_{p1},\; \psi_{a1}^0,\; \psi_{v1}, \;
\psi_{a0}, \; \psi_{a2}, \; \psi_{t1}^0, \; \psi_{t0}, \;
\psi_{t2}. \label{deutcomp} \end{eqnarray} have been computed as
scalar functions of two variables, $p_0$ and $|{\bf p}|$. Now
these components  are available in the form of an analytical
parametrization~\cite{up}.

Since the  one-boson exchange potential~\cite{tjond,tjonnn} was
used   only with a minor adjustment,  our solution does not
contain new physics (or different physics) than the pioneering
paper.  All parameters are presented  in Table 1. The set of
parameters includes the nucleon mass, $m$, the deuteron binding
energy, $\varepsilon_D$, and, therefore, the deuteron mass, $M_D
= 2m+\varepsilon_D$.  The coupling constants are shown in
accordance with our definition of the meson-nucleon form-factors:
\begin{eqnarray} F_B(k) = \frac{\Lambda^2 - \mu_B^2}{\Lambda^2 -
k^2},  \label{formfactor} \end{eqnarray} where $\Lambda$ is a
cut-off parameter (see Table~I).

\subsection{Static properties  and densities} \label{sec:statden}

In order to compare different models of the deuteron and  define
to what degree  each of  the models   is {\em realistic}, it is
natural to calculate various observables within these models and
compare them with each other and with the experimental data. The
canonical way to do this is to calculate the nucleon
contributions  to   static observables, such as mean square
radius, magnetic and quadrupole moments,
etc.~\cite{paris,bonn,gross2,tjond,gross,kuco}. (The mesonic
corrections are quite small and depend upon additional  model
assumptions.) Most of the modern realistic models give quite
good  description of the static properties, or have a decent
explanation in the case  of a minor discrepancy. The observables
of the reactions, such as formfactors and structure functions,
are more model dependent. Those are often used as a tool for
discriminating between   models or certain ways of using models
for calculations.

A direct comparison of the wave functions and amplitudes usually
seems to be less meaningful. Still, in many cases particular
properties of the  wave functions and amplitudes directly
indicate what will happen when observables are calculated.  A
good example here is the $D$-wave admixture, which directly
affects values of the magnetic and quadrupole moments, tensor
analysing power, $T_{20}$, and spin-dependent structure functions
of the deuteron, $g_1^D$ and $b_2^D$. Another example, the high
momentum ``tail'' of the wave function (amplitude) dominates in
the observables for some kinematic  conditions with a high
momentum transfer. However,    the interpretation and properties
of   BS amplitudes are very different from the ones of the wave
functions, a direct  comparison between them is impossible.

Recently, it has been argued that a new intuition can be
developed in  working with the BS amplitude~\cite{kuco,up}. The
approach is based on dealing  with the  charge and current
densities. The same densities also can be calculated in the
nonrelativistic approach with wave functions. Thus we gain a
possibility to compare the models by comparing densities. Besides
densities are more directly linked to the observables than wave
functions and amplitudes. Following this guiding  idea, we
calculate the  charge densities which are related to the various
structure functions in   deep inelastic scattering.

The nucleon contribution to  the deuteron  observable,
$\langle{\cal O}\rangle$, in many important cases is defined by
the triangle diagram (see Fig.~\ref{tre-dia}): \begin{eqnarray}
&&\langle {\cal O} \rangle_M  = i\int \frac{d^4p}{(2\pi)^4} {\sf
Tr}\left\{ \bar\Psi_M(p_0,{\bf  p})\hat {\cal O}(p_1, q, k)
\Psi_M(p_0+k_0,{\bf  p+ k}) (\hat p_2-m) \right \}, \label{me}
\end{eqnarray} where $p_{1,2} = P_D/2\pm p= (M_D/2 \pm p_0, \pm
{\bf  p})$,  $P_D=(M_D , {\bf  0})$ is the deuteron momentum in
the rest frame,  $p = (p_0,{\bf  p})$ is the relative momentum of
nucleons, $\hat {\cal O}(p_1,q,k)$ is an   appropriate operator
and  $\Psi_M(p_0,{\bf  p})$ is the Bethe-Salpeter amplitude for
the deuteron with $M$ being  the deuteron's total angular
momentum projection (see Ref.~\cite{1} for definition and
conventions for  $\Psi_M(p_0,{\bf  p})$).  The operator $\hat
{\cal O}(p_1,q,k)$ is written in a general form, depending on
nucleon momentum, $p_1$ and two external momenta, $q$ which does
not  sneak into the lower part of the diagram and $k$ which adds
to nucleon momentum. In  the present paper we consider   only the
case with  $k = 0$.

The explicit form of operators and structure of the deuteron
matrix elements relevant to the deep inelastic scattering, were
studied   previously~\cite{bls,unon,moff,mf2,mg1,1}. One of the
methods to find a form of the  operator for deep inelastic
scattering is to adopt the Wilson's Operator Product
Expansion~\cite{wilson} and use it within the effective meson
nucleon theory~\cite{bls,unon,cksu,1}. Other possible approach is
based on a parametrization of the operators in the most general
form with analysis of all structures phenomenologically  or
within models for quark-nucleon
amplitudes~\cite{moff,mf2,mg1,msm,grossdis}. Both approaches lead
to the same  results in ``convolution approximation'' which
assumes that the deformation of the nucleon operator
off-mass-shell can be neglected. Very small and strongly model
dependent corrections  to the convolution are qualitatively
similar in both approaches. In the present paper we do not
concentrate on the differences  between  these two approaches,
i.e. our analysis is not going beyond the convolution
approximation.

Thus, the relevant operators are   vector, $\gamma_{\mu}$,  and
axial vector, $\gamma_5\gamma_{\mu}$. Let us define the matrix
elements of the following components of these operators:
\begin{eqnarray} && \left \{ \begin{array}{c} \langle {\gamma_0}
\rangle_M  \\ \langle {\gamma_3} \rangle_M  \\ \langle
{\gamma_5\gamma_0} \rangle_M  \\ \langle {\gamma_5\gamma_3}
\rangle_M   \end{array} \right \} = \frac{ i}{2M_D}\int
\frac{d^4p}{(2\pi)^4} {\sf Tr}\left\{ \bar\Psi_M(p_0,{\bf
p})\right. \left \{ \begin{array}{c} \gamma_0 \\ \gamma_3 \\
\gamma_5\gamma_0 \\ \gamma_5\gamma_3 \end{array} \right \} \left.
\phantom{\bar\Psi}\!\!\!\!\!\! \Psi_M(p_0,{\bf  p}) (\hat p_2-m)
\right \}, \label{meop} \end{eqnarray} where notation in the
l.h.s. are obvious and we use this notation in what follows.
There are three main combinations of these  matrix elements which
are important for deep inelastic scattering and which are
discussed in the present paper. They are presented below together
with the correspondent structure functions of the deuteron:
\begin{eqnarray} F_2^D \quad \to \quad &&\frac{1}{3} \sum_M
\left \{  \langle \gamma_0 \rangle_M + \langle \gamma_3
\rangle_M  \right \} \label{meun}\\ b_2^D \quad \to \quad &&
\left \{ \langle \gamma_0\rangle_{M=1} + \langle \gamma_3
\rangle_{M=1} \right \}   - \left \{ \langle \gamma_0
\rangle_{M=0} + \langle \gamma_3 \rangle_{M=0}\right \}
\label{mete}\\ g_1^D \quad \to \quad && \left \{ \langle \gamma_5
\gamma_0\rangle_{M=1}  +  \langle \gamma_5\gamma_3  \rangle_{M=1}
\right \}, \label{mepo} \end{eqnarray} where $\to$ means
``corresponding to'' and does not have a direct mathematical
interpretation. Explicit connection of these matrix elements to
the structure  functions will be discussed in Section
\ref{sec:th}.  The important circumstance now is that these
matrix elements present  charges and currents, vector and axial
vector, calculated for the deuteron states with different
combinations of the total angular momentum projections, $M$.
Properties of these  matrix elements, including rigorous sum
rules for the charges, are used  later to discuss the structure
functions.

Let us start from the matrix elements involved in
eq.~(\ref{meun}). The first term, $\langle \gamma_0 \rangle_M $,
is the   charge of the conserved vector current  and since it is
independent of  $M$ the following sum rules can be immediately
written down (see e.g.~\cite{bls,1}): \begin{eqnarray}
\frac{1}{3} \sum_M  \langle \gamma_0 \rangle_M  =  \langle
\gamma_0 \rangle_M  = 1. \label{srun} \end{eqnarray} Obviously
this sum rule is also valid in the nonrelativistic approaches,
since it is just a normalization condition for both the  wave
functions and BS amplitudes. To make more meaningful use of this
matrix element and compare    our approach to other realistic
approaches, we define charge density  in the BS formalism as (see
Appendix \ref{sec:dens} for explicit form): \begin{eqnarray}
\langle \gamma_0 \rangle^{BS}  ({\bf p}) = \frac{i}{12\pi^2 M_D}
\sum_M\int \frac{dp_0}{2\pi} {\sf Tr}\left\{ \bar\Psi_M(p_0,{\bf
p})  \gamma_0   \Psi_M(p_0,{\bf  p}) (\hat p_2-m) \right
\},\label{nchbs} \\ \frac{1}{4\pi} \int  {d\bf p}  \langle
\gamma_0 \rangle^{BS} ({\bf p}) = 1 , \label{nchnbs}
\end{eqnarray} which in the nonrelativistic limit corresponds to
the momentum density~\cite{unon,1,kulg1}: \begin{eqnarray}
\langle \gamma_0 \rangle^{NR}  ({\bf p})= \frac{1}{6\pi^2}
\sum_M\left |\Psi_M({\bf p})\right |^2 = \left (u^2(|{\bf p}|) +
w^2(|{\bf p}|)\right ), \label{nchnr} \\ \frac{1}{4\pi}  \int
{d\bf p}  \langle \gamma_0 \rangle^{NR}({\bf p}) = 1,
\label{nchnnr} \end{eqnarray} where $\Psi_M({\bf p})$ is a
nonrelativistic wave function of the deuteron (do not confuse
with the Bethe-Salpeter amplitude, $\Psi_M(p_0, {\bf p})$!) with
$u$ and $w$ being its $S$- and $D$-wave components. The
densities, $\langle \gamma_0 \rangle  ({\bf p})$, calculated in
three realistic models~\cite{paris,bonn,urel} are shown in
Fig.~\ref{g0den} as functions of $|{\bf p}|$.  Certain model
differences at $|{\bf p}| > 0.5$~GeV are present, however there
is no  distinguishing feature of the relativistic density.

Similarly, we consider other components of the vector and axial
matrix elements. The relativistic densities, $\langle \gamma_3
\rangle^{BS}  ({\bf p})$, $\langle \gamma_5\gamma_0\rangle^{BS}
({\bf p})$, etc, are defined by replacing $\gamma_0$  with
$\gamma_3$, $\gamma_5 \gamma_0$, etc, and taking appropriate
combination of the matrix elements with different $M$ in
eq.~(\ref{nchbs}). The explicit expressions for all densities in
terms of   components of the BS amplitude are presented in the
Appendix~\ref{sec:dens}.

The nonrelativistic densities are calculated using methods of
nonrelativistic reduction~\cite{unon,1,kulg1}.  Here we present
the most interesting of them.  A remarkable feature of the
non-relativistic  densities is that they have the same angular
dependences as the corresponding relativistic densities.

The 3rd spatial component of the vector current in the
nonrelativistic limit gives: \begin{eqnarray} \langle \gamma_3
\rangle^{NR}({\bf p})= \frac{p_3}{6\pi^2m} \sum_M\left
|\Psi_M({\bf p})\right |^2   + {\cal O}\left (\frac{|{\bf
p}|^3}{m^3} \right )\simeq \frac{|{\bf p}|cos \vartheta}{m} \left
(u^2(|{\bf p}|) + w^2(|{\bf p}|)\right ) , \label{n3nr}
\end{eqnarray} where $\vartheta$ is polar angle of vector ${\bf
p}$. Note, this density  has a suppression factor of $\sim |{\bf
p}|/m$, due to a mixing of upper and lower components of the
Dirac spinors by non-diagonal matrix $\gamma_3$. Since the
angular dependence of the explicit form of the relativistic
density  $\langle \gamma_3 \rangle^{BS}  ({\bf p})$  is also
absorbed in the  factor $cos\vartheta $ (see Appendix
\ref{sec:dens}), the  integrals of both relativistic and
nonrelativistic densities $\langle \gamma_3 \rangle$ over
${d\bf p}$ are zero: \begin{eqnarray} \int {d\bf p}  \langle
\gamma_3 \rangle^{BS}({\bf p})=  \int {d\bf p}  \langle \gamma_3
\rangle^{NR}({\bf p}) = \langle \gamma_3 \rangle \propto
\int\limits_{-1}^{1} d \left( cos \vartheta \right )cos \vartheta
= 0, \label{n3cos} \end{eqnarray}

These densities for realistic models  are show in
Fig.~\ref{g3den} at $\theta = 0$. Basically they just reflect the
behavior of the charge densities from Fig.~\ref{g0den} in
accordance with the nonrelativistic formula (\ref{n3nr}). For
illustration,   the curve representing the BS density from
Fig.~\ref{g0den} multiplied by the factor $|{\bf p}|/m$ is also
shown  (dash-dotted). Surprisingly enough, it barely can be
distinguished from the exact  $\langle \gamma_3 \rangle^{BS}
({\bf p})$ (solid line) even at momenta higher than $m$.

Two examples of spin-dependent densities:   \begin{eqnarray}
\langle \gamma_0 \rangle^{NR}_{M=1}({\bf p}) -\langle \gamma_0
\rangle^{NR}_{M=0}({\bf p})\simeq -\frac{3}{2} P_2(cos
\vartheta)  w(|{\bf p}|)\left (2\sqrt{2}u(|{\bf p}|) + w(|{\bf
p}|)\right ) , \label{dn0nr} \end{eqnarray} \begin{eqnarray}
\langle\gamma_5 \gamma_3 \rangle^{NR}_{M=1}({\bf p})  \simeq
u^2(|{\bf p}|) - \frac{1}{2}w^2(|{\bf p}|)+ P_2(cos
\vartheta)w(|{\bf p}|) \left (w(|{\bf p}|)-\sqrt{2}u(|{\bf
p}|)\right ) , \label{n53nr} \end{eqnarray} where $P_2(x)$ is the
Legendre polynomial.  We also can easily write down sum rules for
the spin-dependent densities (\ref{dn0nr}) and (\ref{n53nr}):
\begin{eqnarray} \int {d\bf p}  \left\{ \langle \gamma_0
\rangle^{NR}_{M=1}({\bf p}) -\langle \gamma_0
\rangle^{NR}_{M=0}({\bf p})\right \}\propto  \int\limits_{-1}^{1}
d \left( cos \vartheta \right ) P_2(cos \vartheta) = 0,
\label{dnsr} \end{eqnarray} \begin{eqnarray}  \frac{1}{4\pi}\int
{d\bf p}  \langle \gamma_5\gamma_3 \rangle^{NR}({\bf p}) =
1-\frac{3}{2}w_D, \label{naxnr} \end{eqnarray} where $w_D$ is the
weight of the $D$-wave in the deuteron. The relativistic
analogue  of the sum rule (\ref{naxnr}) can be used for an
estimate of the ``admixture'' of the $D$-wave in the relativistic
formalism which otherwise does not allow for probabilistic
interpretation.  Numerically we have:   \begin{eqnarray}
\frac{1}{4\pi}\int {d\bf p}  \langle \gamma_5\gamma_3
\rangle^{BS}({\bf p}) \simeq  0.9215, \label{naxbs}
\end{eqnarray} which gives us an estimate of $w_D\approx
5\%$.\footnote{ There can be corrections, ${\cal O}(|{\bf
p}|^2/m^2)$, to the density, $\langle\gamma_5 \gamma_3
\rangle^{NR}_{M=1}({\bf p})$, see refs.~\cite{cksu,kulg1}.
However, their estimated contribution to integral (\ref{naxnr})
is small,  ${\
\lower-1.2pt\vbox{\hbox{\rlap{$<$}\lower5pt\vbox{\hbox{$\sim$}}}}\
} 1\%$.}

The  realistic model densities (\ref{dn0nr}) and (\ref{n53nr})
are  presented in Figs.~\ref{dg0den} and~\ref{g53den},
respectively. These two examples confirm the conclusions of the
previous  illustrations: (i) realistic models are in a reasonable
agreement among each other,  providing description of the charge
and current densities of the deuteron, and (ii) in spite of some
model variations at high momentum, there is no distinguishing
feature of the densities obtained in a relativistic BS formalism.
Therefore, we can not expect striking relativistic effects caused
by the {\em form} of the   densities.  However, this conclusion
does not close a possibility for effects generated by the
differences in the relativistic and non-relativistic  description
of the deep inelastic reaction.

\section{Relativistic theory of the deep inelastic scattering on
the deuteron} \label{sec:th}

\subsection{Definitions and kinematics} We start with the general
form of the hadronic tensor  of the deuteron with the total
angular momentum projection, $M$, keeping  only  leading twist
structure functions: \begin{eqnarray} W_{\mu\nu}^D(q,P_D,M) =&&
\left ( -g_{\mu\nu} +\frac{q_\mu q_\nu}{q^2}\right )
F_1^D(x_D,Q^2,M) +  \left ( P_{D\mu} - q_\mu \frac{P_Dq}{q^2}
\right )  \left ( P_{D\nu} - q_\nu \frac{P_Dq}{q^2}  \right )
\frac{F_2^D(x_D,Q^2,M)}{P_Dq} \nonumber\\ && +\frac{iM_D}{P_Dq}
\epsilon_{\mu\nu\alpha\beta}  q^\alpha S_D^\beta(M)
g_1^D(x_D,Q^2), \label{htend} \end{eqnarray} where
$q=(\nu,0,0,-\sqrt{\nu^2+Q^2})$ is the momentum transfer, $
Q^2=-q^2$, $x_D = Q^2/(2P_Dq)$ (in the rest frame of the
deuteron  $x_D = Q^2/(2M_D\nu)$),  $S_D(M)$ is the deuteron spin
(see Appendix~\ref{sec:hadron}) and $F_{1,2}^D$ and $g_1^D$ are
the deuteron structure functions.  Averaged over the $M$ this
expression leads to the well-known form of the spin-independent
hadronic tensor which is valid for hadrons with any spin:
\begin{eqnarray} W_{\mu\nu}^D(q,P_D) &=& \frac{1}{3}\sum_M
W_{\mu\nu}^D(q,P_D,M) \label{av}\\ &=& \left ( -g_{\mu\nu}
+\frac{q_\mu q_\nu}{q^2}\right ) F_1^D(x_D,Q^2) +  \left (
P_{D\mu} - q_\mu \frac{P_Dq}{q^2}  \right )  \left ( P_{D\nu} -
q_\nu \frac{P_Dq}{q^2}  \right ) \frac{F_2^D(x_D,Q^2)}{P_Dq},
\label{htenda} \end{eqnarray} where $F_{1,2}^D(x_D,Q^2)$ are
result of averaging of the SF  $F_{1,2}^D(x_D,Q^2,M)$. It is easy
to find that other structure  functions can be  obtained as a
result of other combinations of $W_{\mu\nu}^D(q,P_D,M)$     with
different $M$: \begin{eqnarray} &&  W_{\mu\nu}^D(q,P_D,M=1)  -
W_{\mu\nu}^D(q,P_D,M=-1) \, \propto \, g_1^D(x_D,Q^2),
\label{g1d} \\ && W_{\mu\nu}^D(q,P_D,M=1)-W_{\mu\nu}^D(q,P_D,M=0)
\, \propto \, b_{1,2}^D(x_D,Q^2) \label{b12d} \end{eqnarray} The
eqs. (\ref{av}), (\ref{g1d}) and (\ref{b12d}) are the  basis for
the experimental measurements of the deuteron structure
functions. However, for the purpose of theoretical studies of
the  hadronic tensor  and structure functions the projection
technique is more convenient. All relevant formulae for
projection technique are presented in Appendix~\ref{sec:hadron}.
More background information on the SF $b_{1,2}^D$ can be found in
refs.~\cite{et,fs1,jm,kub}.

To calculate the hadronic tensor of the deuteron we follow  the
general formalism of our approach: \begin{eqnarray}
&&W_{\mu\nu}^D(q,P_D,M)  = i\int \frac{d^4p}{(2\pi)^4} {\sf
Tr}\left\{ \bar\Psi_M(p_0,{\bf  p})\hat {W}^N_{\mu\nu}(p_1,q)
\Psi_M(p_0,{\bf  p}) (\hat p_2-m) \right \}, \label{w}
\end{eqnarray}

The nucleon tensor operator,  $\hat W^N_{\mu\nu}(q,p)$, has been
studied extensively in  recent
years~\cite{bls,cksu,1,moff,mf2,kulg1,msm}. We use a well
established form of the operator, leading to the  convolution
formula~\cite{jaffe}: \begin{eqnarray} &&\hat
W^N_{\mu\nu}(q,p)=\hat W_{\{\mu\nu\}}(q,p) +  \hat
W_{[\mu\nu]}(q,p) \label{deftens}\\[3mm] &&\hat
W_{\{\mu\nu\}}(q,p) = \frac{\hat q}{2pq}W^N_{\mu\nu}(q,p),
\label{ops}\\ &&\hat W_{[\mu\nu]}(q,p) =  \frac{ i}{2pq}
\epsilon_{\mu\nu\alpha\beta} q^{\alpha} \gamma^\beta \gamma_5
g_1^N (q,p) , \label{opa} \end{eqnarray} where $\{\ldots \}$ and
$[\ldots ]$ denote symmetrization and  antisymmetrization of
indices, respectively, and $g_1^N(q,p)=g_1^N(x,Q^2)$ is the
spin-dependent nucleon  SF.   The hadronic  tensor of the
nucleon, $W_{\mu\nu}(p,q)^N$, is defined as:  \begin{eqnarray}
W_{\mu\nu}^N(q,p) =  \left ( -g_{\mu\nu} +\frac{q_\mu
q_\nu}{q^2}\right ) F_1^N(x,Q^2) +  \left ( p_\mu - q_\mu
\frac{pq}{q^2}  \right )  \left ( p_\nu - q_\nu \frac{pq}{q^2}
\right ) \frac{F_2^N(x,Q^2)}{pq}, \label{hten}\end{eqnarray}
where $x = Q^2/(2pq)$ and $F_{1,2}^N$  are the nucleon  SFs. The
small effects of the off-mass-shell deformation of the nucleon
tensor\cite{moff,kulg1,uoff} are not considered here, since these
effects do not affect sum rules for SF  and do not noticeably
change the absolute values of the  SFs. That is why  SFs
$F_{1,2}^N$ in (\ref{hten}) do not depend on $p^2$, but  $q^2$
and $pq$.

Using  projectors (Appendix~\ref{sec:hadron}), we extract   SFs
from the hadronic tensor of the deuteron: \begin{eqnarray} &&
F_1^D(x_N,Q^2,M) = i\int \frac{d^4p}{(2\pi)^4} {F_1^N\left(
\frac{x_N m}{p_{10}+p_{13}}, Q^2\right)}  \frac{ {\sf Tr}\left\{
\bar\Psi_M(p_0,{\bf p})(\gamma_0+\gamma_3) \Psi_M(p_0,{\bf p})
(\hat p_2-m) \right \}}{2(p_{10}+p_{13})}, \label{f1m}\\
&&F_2^D(x_N,Q^2,M) = i\int \frac{d^4p}{(2\pi)^4} {F_2^N} \left(
\frac{x_N m}{p_{10}+p_{13}}, Q^2\right)  \frac{ {\sf Tr}\left\{
\bar\Psi_M(p_0,{\bf p})(\gamma_0+\gamma_3) \Psi_M(p_0,{\bf p})
(\hat p_2-m) \right \}}{2M_D}, \label{f2m}\\ &&g_1^D(x_N,Q^2) = i
\int \frac{d^4p}{(2\pi)^4} {g_1^N}\left( \frac{x_N
m}{p_{10}+p_{13}}, Q^2\right)  \frac{\left. {\sf Tr}\left\{
\bar\Psi_M(p_0,{\bf p})(\gamma_0+\gamma_3)\gamma_5
\Psi_M(p_0,{\bf p}) (\hat p_2-m) \right \}\right
|_{M=1}}{2(p_{10}+p_{13})}, \label{g1m} \end{eqnarray} where
$x_N=Q^2/(2m\nu)$ is the Bjorken scaling  variable\footnote{Note
that the ``native'' deuteron variable is $x_D = (m/M_D)x_N$,
however $x_N$ is used more often.},  i.e. this is $x$  for the
on-mass-shell nucleon at rest, $p_{10}$ and $p_{13}$ are the time
and 3-rd components of the struck nucleon momentum.  Formulas
(\ref{f1m}) and (\ref{f2m}) have not been averaged over the
projection  $M$  since in the present form it gives an
understanding of the  SF $b_{1,2}^D$. Indeed, for instance
eq.~(\ref{f2m}) gives two independent ``SFs'', with $M = \pm 1$
and $M=0$, which are related to the usual spin-independent  SF,
$F_2^D$, and a new  SF, $b_2^D$: \begin{eqnarray}
&&F_2^D(x_N,Q^2) = \frac{1}{3} \sum_{M=0,\pm 1} F_2^D(x_N,Q^2,M),
\label{f2}\\[1mm] && b_2(x_N,Q^2) =
F_2^D(x,Q^2,M=+1)-F_2^D(x,Q^2,M=0),\label{b2}    \\[3mm] &&
F_2^D(x_N,Q^2,M=+1) = F_2^D(x_N,Q^2,M=-1).\label{pm}
\end{eqnarray}

Note, the SF $F_2^D(x,Q^2,M)$ is independent of the  lepton
polarization, therefore, both  SFs, $F_2^D$  and $b_2^D$, can be
measured in  experiments with an unpolarized lepton beam and
polarized deuteron target. In view of  eq.~(\ref{pm}), only one
of the  SFs $F_2^D(x,Q^2,M)$ is needed, in addition to the
spin-independent $F_2^D(x,Q^2)$, in order to obtain $b_2(x,Q^2)$.
The other  SF, $b_1^D$, is related to the deuteron  SF $F_1^D$,
the same way as $b_2^D$ is related to $F_2^D$, viz. via
eqs.~(\ref{f2}), and $b_2^D = 2xb_1^D$.

\subsection{ Singularities of the triangle diagram and
calculation of structure functions}

It has been  shown previously~\cite{fs,land,km,msm}  how a
singular structure of the triangle graph (Fig.~\ref{tre-dia})
rules the behavior of the spin-independent   SF $F_2^D$. In
particular, it is found that the relativistic impulse
approximation satisfies the unitarity and provides the correct
kinematical region of the variable $x_N$. However,  both  these
properties of the exact covariant amplitude are violated in
practical  calculations, when  nonrelativistic wave functions of
the deuteron  are used. In this case  one can refer to the
argument that such deviations are small, and  are not important
for phenomenology. At the same time, a realistic Bethe-Salpeter
amplitude of the deuteron serves ideally for a  consistent
phenomenological application of the covariant theory of the
processes on the bound nucleons.

In order to calculate  SFs, (\ref{f2m})-(\ref{b2}) and analyze
the sum rules, the singularities of the triangle diagram should
be explicitly taken into account. To do that, eq.~(\ref{f2m}) is
rewritten as: \begin{eqnarray} &&F_2^D(x_N,Q^2,M)   =
\frac{i}{2M_D}\int \frac{d^4p}{(2\pi)^4}  F_2^N \left( \frac{x_N
m}{p_{10}+p_{13}}, Q^2\right) \nonumber \\[1mm] && \quad\quad
\frac{1}{(p_1^2-m^2+i\epsilon)^2(p_2^2-m^2+i\epsilon)}  {\sf
Tr}\left\{ \bar\phi_M(p_0,{\bf p}) (\hat p_1+m)
(\gamma_0+\gamma_3) (\hat p_1+m) \phi_M(p_0,{\bf p}) (\hat p_2+m)
\right \}, \label{sing1} \end{eqnarray} where $\phi_M(p_0,{\bf
p}) = (\hat p_1-m)\Psi_M(p_0,{\bf p})(\hat p_2-m)$ is the
Bethe-Salpeter vertex function of the deuteron.

Analysis of singularities in the complex $p_{2+}$-plane allows
for one  analytical  integration in (\ref{sing1})~\cite{km}.
After translation into  variables which are used in the present
paper, this integration is equivalent to picking  the residue of
the second nucleon pole, $p_{20} = \omega  = \sqrt{m^2+{\bf
p}^2}$ or  $p_0 = M_D/2 - \omega $, in the complex $p_{0}$-plane
when both of  the following conditions are satisfied:
\begin{eqnarray} 0 < \omega  - p_{3} < M_D. \label{sing2}
\end{eqnarray} The contribution of the region of $p$  beyond
(\ref{sing2}), to the integral (\ref{sing1}), is zero, i.e.
different poles cancel each other.  Note,  that  $p_{10} = M_D -
\omega $ in the required pole. Calculating residue in
(\ref{sing1}), one gets:  \begin{eqnarray} &&F_2^D(x_N,Q^2,M) =
\frac{1}{2M_D}\int \frac{d^3{\bf p}}{(2\pi)^3} F_2^N \left(
\frac{x_N m}{M_D- \omega  +p_{3}}, Q^2\right) \Theta(M_D-\omega
+ p_{3} )\label{sing3} \\ && \quad\quad \frac{1}{2\omega M_D^2
(M_D-2\omega  )^2} {\sf Tr}\left\{ \bar\phi_M(p_0,{\bf p}) (\hat
p_1+m)(\gamma_0+\gamma_3) (\hat p_1+m) \phi_M(p_0,{\bf p}) (\hat
p_2+m) \right \}_{p_0 = \frac{M_D}{2}-\omega }, \nonumber
\end{eqnarray} where the $\Theta$-function guaranties the right
of  conditions (\ref{sing2}) is satisfied, and the left condition
is always satisfied.

It is useful to rewrite (\ref{sing3}) in the convolution form:
\begin{eqnarray} && F_2^D(x_N,Q^2,M) = \int\limits_{x_N}^{M_D/m}
dy   F_2^N \left( \frac{x_N}{y}, Q^2\right) f^{N/D}_M(y),
\label{f2ms} \end{eqnarray} where ``the effective distribution''
of nucleons in the deuteron is defined by \begin{eqnarray} &&
f_M^{N/D}(y)   =  \frac{1}{2M_D}\int \frac{d^3{\bf p}}{(2\pi)^3}
\delta\left (  y - \frac{M_D-w+p_3}{m} \right) \Theta(y)
\frac{1}{2\omega M_D^2 (M_D-2\omega  )^2} \label{fndm}\\[2mm] &&
\quad \quad {\sf Tr}\left\{ \bar\phi_M(p_0,{\bf p}) (\hat
p_1+m)(\gamma_0+\gamma_3) (\hat p_1+m) \phi_M(p_0,{\bf p}) (\hat
p_2+m) \right \}_{p_0 = \frac{M_D}{2}-\omega }. \nonumber
\end{eqnarray}

The  SFs $F_{1,2}^D$ and $b_{1,2}^D$ are now  calculated as
follows: \begin{eqnarray} \left \{ \begin{array}{c}
F_1^D(x_N,Q^2) \\ b_1^D(x_N,Q^2) \end{array}\right \} =
\int\limits_{x_N}^{M_D/m}\frac{dy}{y} \left \{ \begin{array}{c}
f^{N/D}(y) \\ \Delta f^{N/D}(y)  \end{array}\right \}  F_1^N
\left( \frac{x_N}{y}, Q^2\right) ,\label{fb1} \\[3mm] \left \{
\begin{array}{c} F_2^D(x_N,Q^2) \\ b_2^D(x_N,Q^2)
\end{array}\right \} = \int\limits_{x_N}^{M_D/m} dy \left \{
\begin{array}{c} f^{N/D}(y) \\ \Delta f^{N/D}(y)
\end{array}\right \}  F_2^N \left( \frac{x_N}{y}, Q^2\right) ,
\label{fb2}  \end{eqnarray} where distributions $f^{N/D}$ and
$\Delta f^{N/D}$ are given by \begin{eqnarray} f^{N/D} (y) &=&
\frac{1}{3}\sum_M f_M^{N/D}(y), \label{fnd} \\ \Delta f^{N/D} (y)
&=& f^{N/D}_1 (y) - f^{N/D}_0(y). \label{dfnd} \end{eqnarray}

Similarly, for the SF $g_1^D$ we get: \begin{eqnarray} &&
g_1^D(x_N,Q^2) = \int\limits_{x_N}^{M_D/m}  \frac{dy}{y}   g_1^N
\left( \frac{x_N}{y}, Q^2\right) \vec f^{N/D}(y), \label{g1ms}
\end{eqnarray} where the effective polarized distribution of
nucleons in the deuteron is defined by \begin{eqnarray} && \vec
f^{N/D}(y)   =  \frac{1}{2M_D} \int \frac{d^3{\bf p}}{(2\pi)^3}
\delta\left (  y - \frac{M_D-w+p_3}{m} \right) \Theta(y)
\frac{1}{2\omega M_D^2 (M_D-2\omega  )^2} \label{fndpol}\\[2mm]
&&  \quad \quad {\sf Tr}\left\{ \bar\phi_{M=1}(p_0,{\bf p}) (\hat
p_1+m)(\gamma_0+\gamma_3) \gamma_5(\hat p_1+m)
\phi_{M=1}(p_0,{\bf p}) (\hat p_2+m) \right \}_{ p_0 =
\frac{M_D}{2}-\omega} \nonumber \end{eqnarray}

\subsection{Sum rules for the deuteron structure functions}

Two sum rules can be written   for the effective distribution
$f^{N/D}_M(y)$: \begin{eqnarray} &&\int\limits_{0}^{M_D/m}
f^{N/D}_M (y) dy= \frac{1}{4\pi}\int d{\bf p} \langle\gamma_0
\rangle_M ({\bf p})= \langle\gamma_0 \rangle_M = 1, \label{ch0}
\\ &&\int\limits_{0}^{M_D/m} y f^{N/D}_M (y) dy= \langle D|
\left (\Theta_N \right)_\mu^\mu  |D \rangle_M = 1-\delta_N,
\label{emt} \end{eqnarray} where   $ \left
(\Theta_N\right)_\mu^\mu \propto i \bar
\psi(x)\gamma_\mu\partial^\mu\psi(x)$ is the trace of the
energy-momentum tensor.  Eq.~(\ref{ch0})  presents    the vector
charge conservation generalized for the   deuteron states with
different $M$ (see eqs.~(\ref{nchbs})-(\ref{nchnbs})). In spite
of  clear physical interpretation,  some time ago it was a
subject of some controversy~\cite{fs,land,km}. Indeed, the
derivation of sum rule (\ref{ch0}) contains some subtle points
and equivalence between it and the expression for the charge,
(\ref{srun}), is a non-trivial fact, particularly, because of the
presence of the $\Theta$-function in eq.~(\ref{fndm}). This
$\Theta$-function provides  correct kinematics in  variable
$x_N$  but cuts out a part  of the integration domain in $d^3{\bf
p}$.  This cutting of the integration interval in the polar angle
$\theta$ leads to non-zero   contribution of the  matrix element
containing $\gamma_3$, which is proportional to $cos\theta$.
However, the validity of this sum rule has been  firmly
established~\cite{fs,land,km}. The   sum rule (\ref{emt}) for the
first moment of $f_M^{N/D}$ is of a different nature, it presents
the nucleon contribution  to the total momentum of the
deuteron~\cite{bls,kulmes,unon,1} and $\delta_N$ is a part of the
total momentum carried by the non-nucleon component (mesons). The
constant $\delta_N$ cam not be fixed  in a model independent
fashion, rather it is calculated within  any particular model.
Self-consistency of the theory requires that meson exchange
current contribution into the deuteron SF, $F_2^D$, exactly
compensates the loss of energy by nucleons, (\ref{emt}). An {\em
important} property of sum rules, (\ref{ch0}) and (\ref{emt}), is
that their r.h.s. does not depend on the deuteron spin
orientation.

The sum rules for  $f^{N/D}(y)$  and $\Delta f^{N/D}(y)$ follow
from sum rules for  $f^{N/D}_M(y)$ and definitions
(\ref{fnd})-(\ref{dfnd}): \begin{eqnarray}
&&\int\limits_{0}^{M_D/m} f^{N/D} (y) dy  =  \frac{1}{3}\sum_M
\langle \gamma_0 \rangle_M =\langle \gamma_0 \rangle= 1,
\label{ch1}\\ && \int\limits_{0}^{M_D/m} y f^{N/D} (y) dy =
\frac{1}{3}\sum_M \langle D|   \left (\Theta_N \right)_\mu^\mu |D
\rangle_M  = 1-\delta_{N}, \label{emt1}\\
&&\int\limits_{0}^{M_D/m}\Delta f^{N/D} (y) dy  =  \langle
\gamma_0 \rangle_{M=1} - \langle \gamma_0 \rangle_{M=0}=0,
\label{ch2}\\ && \int\limits_{0}^{M_D/m} y \Delta f^{N/D} (y) dy
= \langle D|   \left (\Theta_N \right)_\mu^\mu |D \rangle_{M=1} -
\langle D|   \left (\Theta_N \right)_\mu^\mu |D \rangle_{M=0} =
0. \label{emt2} \end{eqnarray}

Sum rules for the deuteron  SFs $b_1^D$ and $b_2^D$ are the
immediate result of combining eqs.~(\ref{ch2})-(\ref{emt2}) and
(\ref{fb1})-(\ref{fb2}): \begin{eqnarray} \int\limits_0^1 dx_D
b_1^D(x_D) =0, \label{sr1}\\ \int\limits_0^1 dx_D b_2^D(x_D) =0,
\label{sr2} \end{eqnarray} in an agreement with the sum rules
suggested by Efremov and Teryaev~\cite{et}.

Sum rule for the spin-dependent distribution relates the integral
of the spin-dependent distribution of nucleons to the 3-rd
component of the axial current, (\ref{naxbs}): \begin{eqnarray}
&&\int\limits_{0}^{M_D/m} \vec f^{N/D} (y) dy  =  \langle
\gamma_5\gamma_3 \rangle_{M=1}^{BS}. \label{axc} \end{eqnarray}

An explicit expression  for the distribution  function
$f^{N/D}_M(y)$ (and therefore of $f^{N/D}(y)$ and  $\Delta
f^{N/D}(y)$) in terms of the components of the  Bethe-Salpeter
amplitude can be directly established from eqs.  (\ref{fndm}) and
(\ref{fndpol})  and formulae for the  corresponding densities are
given in Appendix~\ref{sec:dens}.

\subsection{Calculation of distributions: Inverse Wick rotation.}

To calculate numerically the effective distribution functions,
(\ref{fndm}), (\ref{fnd}), (\ref{dfnd}) and (\ref{fndpol}), we
need to know  the matrix elements over the BS vertex functions as
functions of ${\bf p}$ and at $p_0 = M_D/2-\sqrt{m^2+{\bf p}^2}$
along real $p_0$. Let us remind that the components of BS
amplitude have been found along imaginary axis in $p_0$-plane
(Wick rotation).

We explain the procedure of calculating the ``inverse Wick
rotated'' matrix element using an example of $f^{N/D}(y)$. Other
distributions  are calculated the same way. First, let us rewrite
eq.~(\ref{fndm}) in the form: \begin{eqnarray} f^{N/D}(y)   =
\frac{1}{2M_D}\int \frac{d^3{\bf p}}{(2\pi)^3}  \delta\left (  y
- \frac{M_D-w+p_3}{m} \right) \Theta(y) \left (\langle
\gamma_0\rangle^{BS}_{pole}({\bf p}) +\langle\gamma_3
\rangle^{BS}_{pole}({\bf p}) \right ), \label{fndmpole}
\end{eqnarray} where $\langle  \gamma_0\rangle^{BS}_{pole}({\bf
p})$ and $\langle\gamma_3 \rangle^{BS}_{pole}({\bf p}) $ are the
densities defined similarly to   eq.~(\ref{nchbs}), but in the
``pole approximation'', i.e. exact integral is replaced by
residue at the second nucleon pole. An explicit form of $
\langle \gamma_0\rangle^{BS}_{pole}({\bf p})$ and
$\langle\gamma_3 \rangle^{BS}_{pole}({\bf p})  $ is clear from
eqs.~(\ref{fndm}) and (\ref{fndmpole}). Second, we notice that
this density does not have any singularities in the complex plane
of $p_0$. Neither does the matrix element: \begin{eqnarray} {\sf
Tr}\left\{ \bar\phi_M(p_0,{\bf p}) (\hat p_1+m)
(\gamma_0+\gamma_3) (\hat p_1+m) \phi_M(p_0,{\bf p}) (\hat p_2+m)
\right \}\nonumber\\ = {(p_1^2-m^2)^2(p_2^2-m^2)} {\sf Tr}\left\{
\bar\Psi_M(p_0,{\bf p})(\gamma_0+\gamma_3) \Psi_M(p_0,{\bf p})
(\hat p_2-m)\right \}. \label{rw1} \end{eqnarray} Therefore, to
calculate the density in the pole approximation, the matrix
element  (\ref{rw1})  can be safely expanded into a Taylor series
in the variable $p_0$ around the point $p_0 = 0$.  Third, the
coefficients of the Taylor expansion can be calculated, using the
known r.h.s. of  eq.~(\ref{rw1}) along imaginary $p_0$. And last,
a convergence of the expansion can be checked numerically, by
comparing results of calculations up to different order in $p_0$.
Note, that eq.~(\ref{rw1}) and similar expressions for all other
distributions are even functions  of $p_0$, therefore the Taylor
expansion really should be done in $p_0^2$. In addition, the
point $p_0 = 0$ should be a good point for the Taylor expansion,
since in the most physically interesting region of $|{\bf p}|/m
\ll  1$ we have $\left | p_0/m \right | \approx \left
|\varepsilon_D/(2m) - |{\bf p}|^2/(2m^2) \right | \ll 1$. The
critical point for any expansion in the nuclear physics is
usually $|{\bf p}|/m =  1$. However, even at this  point $\left |
p_0/m \right |$ is still a good parameter for  the Taylor
expansion,   $\left | p_0/m \right | \approx \left |
\varepsilon_D/(2m) - (\sqrt{2}-1)\right | \sim 0.4$. Parameter
$\left | p_0/m \right |$ is getting close to $1$, with $|{\bf
p}|/m \sim 1.5$.

The results of calculations of the charge density in  the pole
approximation, $\langle \gamma_0\rangle _{pole}({\bf p})$, are
presented in Fig.~\ref{g0pole}, where we compare curves for
calculations up to $\sim p_0^0$ , $\sim p_0^2$ and $\sim p_0^4$.
As we  expected, the procedure is nicely convergent up to $|{\bf
p}|\sim 1$~GeV and with reasonable accuracy can be used up to
$|{\bf p}|\sim 1.5$~GeV. Similar results for the axial density in
pole approximation, $\langle \gamma_5\gamma_3 \rangle
_{pole}({\bf p})$, are shown in Fig.~\ref{g53pole}.

Note, the numerical approximation made in this section,  such as
limiting number of terms in the Taylor expansion, can potentially
cause a violation of the exact sum rules.

\subsection{Nonrelativistic formulae for structure functions}

The nonrelativistic expressions for  $f^{N/D}(y)$,  $\Delta
f^{N/D}(y)$ and $\vec f^{N/D}(y)$  can be obtained by using an
analogy of the charge densities calculated within the
Bethe-Salpeter  formalism  and corresponding densities calculated
with wave functions (e.g. (\ref{nchnr}), (\ref{n3nr}),
(\ref{dn0nr}) and (\ref{n53nr})). Actually, the distributions in
the BS formalism are expressed in  terms of densities in the
nucleon pole approximation, (\ref{fndm}), (\ref{dfnd}) and
(\ref{fndpol}) and not exact densities as (\ref{nchbs}). However,
one can hope that the nucleon pole contribution gives  the main
contribution to the density, at least in the non-relativistic
region. Such type of assumption is very common for nuclear
physics  (see e.g.~\cite{gross2,gross,shakin,mf2,mg1}).  For
instance, the well-known result for the  spin-independent
distribution is immediately reproduced (see
e.g.~\cite{unon,msm,bls}): \begin{eqnarray} f^{N/D}_{NR} (y) &=&
\int \frac{d^3{\bf p}}{(2\pi)^3}  \delta\left
(y-\frac{M_D-w+p_{3}}{m} \right )  \Theta \left (y \right )
\left \{ \langle\gamma_0 \rangle^{NR}({\bf p}) + \langle\gamma_3
\rangle^{NR} ({\bf p})\right \} \nonumber \\ &=&   \int
\frac{d^3{\bf p}}{(2\pi)^3}  \delta\left (y-\frac{M_D-w+p_{3}}{m}
\right )  \Theta \left (y \right )  (1+\frac{|{\bf
p}|cos\theta}{m}) \left \{  u^2(|{\bf p}|) + w^2(|{\bf p}|)
\right \}. \label{fndnr} \end{eqnarray} The presence of the
$\Theta$-function on the r.h.s. of eq.~(\ref{fndnr}) slightly
violates the sum rule (\ref{ch1}).  However, this  is not
noticeable phenomenologically, since the only region of large
momenta, $|{\bf p}| > 0.7$~GeV, is affected by the
$\Theta$-function and it does not  contribute much to the norm of
the deuteron wave function.  We can accept such slight effect of
the $\Theta$-functions, since  a nonrelativistic approximation is
based on the belief that high momenta are not important and if
something is wrong there we just ignore it.

For distribution $\Delta f^{N/D}(y)$, we get: \begin{eqnarray}
\Delta f^{N/D}_{NR} (y) &=& -    \int \frac{d^3{\bf
p}}{(2\pi)^3}  \delta\left (y-\frac{M_D-w+p_{3}}{m} \right )
\Theta \left (y \right ) \label{dfndnr}\\ && \quad\quad\quad
(1+\frac{|{\bf p|}cos\theta}{m}) P_2(cos\theta) \frac{3}{2}
w(|{\bf p}|) \left \{ 2\sqrt{2}u(|{\bf p}|) + w(|{\bf p}|) \right
\}. \nonumber \end{eqnarray} Again, the  sum rule (\ref{ch2}) is
broken by the presence of the $\Theta$-function in
(\ref{dfndnr}). Neglecting it, one gets \begin{eqnarray}
\int\limits_0^1 dx_D b_1^D(x_D)&\propto & \int\limits_0^{M_D/m}
\Delta f^{N/D}_{NR} (y) dy\nonumber\\ & \propto&
\int\limits_{-1}^{1} d(cos\theta) (1+\frac{|{\bf
p|}cos\theta}{m}) P_2(cos\theta) =0, \label{sr1nr} \end{eqnarray}
where the orthogonality property of the Legendre polynomials is
used.

A deviation from zero, caused by the  $\Theta$-functions is not
large compared to 1, but {\em anything} is large compare to 0!
One can {\em artificially} adjust formula (\ref{dfndnr}) to
fulfill this sum rule. For instance, {\em small} corrections to
the normalization of  both terms with $M=1$ and $M=0$ can be made
to satisfy the sum rule in the form (\ref{ch2}). However, the
situation with the second  sum rule, (\ref{sr2}) and
(\ref{sr2nr}), is more difficult and  can not be fixed by any
simple adjustments of the normalizations or ignoring the $\Theta$
functions.  Similar to  eq.~(\ref{sr1nr}), it can be written
(neglecting the  $\Theta$-function!) as: \begin{eqnarray}
&&\int\limits_0^1 dx_D b_2^D(x_D)\propto  \int\limits_0^{M_D/m}
y  \Delta f^{N/D}_{NR} (y) dy\nonumber\\ \quad \quad & \propto&
\int\limits_{-1}^{1} d(cos\theta)  {(M_D-w+|{\bf p|}cos\theta) }
(1+\frac{|{\bf p|}cos\theta}{m}) P_2(cos\theta) \neq 0.
\label{sr2nr} \end{eqnarray} Thus, there is no reason for this
sum rule to be satisfied with the nonrelativistic distribution
function (\ref{dfndnr}). Therefore the nonrelativistic formulae,
in principle, violates the sum rules for the  SFs $b_{1,2}^D$.
However, we can  still hope that it will be a small effect, one
not noticeable in practice.

Nonrelativistic formulae for other spin-dependent distribution,
$\vec f^{N/D}$, are also a straightforward  result of using
density (\ref{n53nr}) instead of the relativistic density in the
pole approximation in eq.~(\ref{fndpol})  (see also footnote to
the formula~(\ref{naxbs})):  \begin{eqnarray} \vec f^{N/D}_{NR}
(y) &=&   \int \frac{d^3{\bf p}}{(2\pi)^3}  \delta\left
(y-\frac{M_D-w+p_{3}}{m} \right )  \Theta \left (y \right )
\label{fndnrpol}\\ && \quad\quad\quad  \left \{ u^2(|{\bf p}|) -
\frac{1}{2}w^2(|{\bf p}|)+ P_2(cos \vartheta)w(|{\bf p}|) \left
(w(|{\bf p}|)-\sqrt{2}u(|{\bf p}|)\right ) \right \}. \nonumber
\end{eqnarray} and the sum rule follows from eq.~(\ref{naxnr}):
\begin{eqnarray} \int\limits_0^{M_D/m}    \vec f^{N/D}_{NR} (y)
dy = 1 - \frac{3}{2}w_D \label{sr53nr} \end{eqnarray}

\subsection{Calculation of distributions: relativistic vs.
nonrelativistic} \label{sec:relnon}

In order to understand if relativistic distribution functions,
(\ref{fndm}), (\ref{dfnd}) and (\ref{fndpol}), are significantly
different from the nonrelativistic distributions, (\ref{fndnr}),
(\ref{dfndnr}) and (\ref{fndpol}), we have to  understand fully
an effect of the one pole approximation on the densities in the
BS formalism.  Indeed, discussion of  Section~\ref{sec:statden}
prepared us to the fact that we can not expect significant
physical effects from the form of the densities, since both
relativistic and nonrelativistic {\em realistic} models lead to
similar results and we do not detect any special trend for  the
relativistic model.

In Figs.~\ref{g0dis} and \ref{g53dis} we compare  the exact
densities and densities in the one pole approximation, for two
most important densities $\langle\gamma_0 \rangle$ and
$\langle\gamma_5\gamma_3 \rangle$.  We find that one pole
approximation leads to a significant change in the density. The
densities in one pole approximation (solid curves) have much
harder tail compared to the exact densities (dotted curves),
starting at  $|{\bf p}| \sim 0.5 m$, and effect is an order of
magnitude at $|{\bf p}| \sim 1.5 m$. This can be qualitatively
understood, considering an example of the charge density,
$\langle\gamma_0 \rangle$. Indeed, picking the nucleon pole in
the full integral, corresponds to neglecting the antinucleon
(negative) contribution to total charge density which
concentrates at high momenta, $|{\bf p}|$. The  presence of
$\Theta$-functions in the expressions for the distribution
functions, $f^{N/D}$ and $\vec f^{N/D}$, cuts off a part of the
high momentum region, but  this is a minor effect. Effect of
``softening'' caused by the $\Theta$-functions is also
illustrated in Figs.~\ref{g0dis} and \ref{g53dis} (dashed lines).
These curves are the BS densities in the pole approximation
(solid lines) multiplied by   $\Theta(M_D-\omega +p_{3})$
integrated over $cos \theta$.

The results presented in Figs.~\ref{g0dis} and \ref{g53dis} imply
that the relativistic densities appearing in the formulae for the
effective distribution functions are enhanced at medium and high
momenta. Comparing these results with those in Figs.~\ref{g0den}
and \ref{g53den} we find that the effect is much larger than any
model differences.  Still, since the effect is concentrated at
high momenta it is not clear if it leads to observable effects in
deep inelastic scattering. To clarify this, in Fig.~\ref{fndp} we
present  the effective distribution function, $f^{N/D}(y)$.  For
completeness of the illustration, we compare  distribution
functions calculated  using three different charge densities: (i)
nonrelativistic  density of the Bonn potential (dotted line),
(ii) relativistic BS density in the one pole  approximation
(solid curve) and (iii) exact densities within the BS approach
(dashed curve). The last curve is aimed to illustrate differences
in the description  of  the {\em mechanism} of   the reaction in
nonrelativistic and relativistic approaches, i.e. it presents
result for made up situation, if charge density of the
nonrelativistic model is  exactly the same as of the BS approach.
We find that consistent relativistic description gives the
effective distribution which is systematically harder at high
momentum fraction, $y$. This is a result  of harder charge
density in the one pole approximation. It is interesting that the
relativistic  distribution  is also enhanced at small $y\to 0$.
This effectively corresponds to larger ``binding effects''  in
the BS approach which was observed in ref.~\cite{urel}. Very
similar effects can be observed in other  effective distribution
functions, $\vec f^{N/D}$ and $\Delta f^{N/D}$. However,  in
these cases those effects are not so explicit, because of the
oscillating nature of the  distribution functions. (See e.g.
discussion about $\Delta f^{N/D}$ below.)

\section{Numerical calculation of structure functions}
\label{sec:num}

\subsection{Unpolarized deuteron, $F_2^D$}

The spin-independent SF of the deuteron, $F_2^D$ is calculated
using  the effective distribution functions  presented in
Fig.~\ref{fndp}  (see discussion at the end of the
Section~\ref{sec:relnon}).  The results of this calculation are
shown in Fig.~\ref{rf2} in the form of  ratio of the SF  of the
deuteron and the nucleon. The nucleon SF, $F_2^N$ is taken from
ref.~\cite{amb} at $Q^2 = 10$~GeV$^2$. We find that BS approach
gives a  behavior of the deuteron SF qualitatively similar to the
results of the nonrelativistic and ``made up nonrelativistic''
calculations.  However, there are two delicate, but essential
differences: \begin{itemize} \item First, the ratio $F_2^D/F_2^N$
in the BS approach  is lower  than others at small $x$, $x< 0.5$.
This effect can be easily understood from  the form of
distribution, $f^{N/D}(y)$, in Fig.~\ref{fndp} and formula
(\ref{fb2}). Indeed, at $x=0$ the SF $F_2^D$ is: \begin{eqnarray}
F_2^D(0) = F_2^N(0) \cdot \int\limits_0^{M_D/m} f^{N/D}(y) dy .
\label{f20} \end{eqnarray} Therefore, $F_2^D(0)/F_2^N(0) =1$
because  $\int\limits_0^{M_D/m} f^{N/D}(y) dy = 1$ is a
normalization, (\ref{ch1}). When we move from $x=0$ to larger
$x$, we ``lose'' part of the  normalization, since the lower
limit of integral in eq.~(\ref{fb2}) is  $x$. Since the
relativistic $f^{N/D}$ is  larger at small $x$ than  the
nonrelativistic  ones, it leads to a faster decrease of the
relativistic  SF with increasing $x$.  The fact that the ratio
$F_2^D/F_2^N$ is less than 1 at small and intermediate $x$ is
know to be a result of the  ``binding of nucleons'' (see
e.g.~\cite{akv,bls,unon,1} and references therein).

\item Second, the relativistic  SF displays sharper rise at
higher $x$, $x>0.5$ than the nonrelativistic ones. This, again,
can be understood from the form of  the distribution,
Fig.~\ref{fndp}, and the convolution formula, (\ref{fb2}).  With
increasing of $x$ the role of the high momentum tail of the
effective distribution is getting more important for integration
in eq.~(\ref{fb2}) and at $x> 1.0$ the  tail is completely
dominating. The deuteron SF at $x>1$ is presented in
Fig.~\ref{f2p}. \end{itemize}

It has been shown recently, that relativistic calculations lead
to the larger binding effects than the nonrelativistic
ones~\cite{tokarev,urel}.  However, the result of
ref.~\cite{tokarev} which presents effects two to three  times
larger than ours  still is neither explained nor  physically
understood. In  ref.~\cite{urel} the size of the effect was not
so large, but the method  of numerical calculations was
essentially based on the  {\em numerical} inverse Mellin
transform of the  non-analytical function. The approximations had
to be made, which eventually was not good at high momentum.  That
is why we specially attempted to verify quantitative size of the
effect. It is worth to remember that we discuss   quite tiny
effect of $\sim 1-2 \%$ in the ratio $F_2^D/F_2^N$. We also can
measure the binding effect using the energy-momentum  sum rule
(\ref{emt}). The quantity $\delta_N$ is a natural parameter
controlling the binding in any particular calculations. For
example, the nonrelativistic formulae allow for an analytical
estimate of $\delta_N$ which essentially gives us an idea about
size of the effect~\cite{bls,unon,urel}: \begin{eqnarray} \delta
=  \frac{\varepsilon_D}{m}  - \frac{\langle T \rangle} {6m} ,
\label{dnon} \end{eqnarray} where $\langle T \rangle$ is a
nonrelativistic mean kinetic energy of nucleons  in the deuteron.
For the realistic models typically  $\langle T \rangle \approx
15$~MeV, which gives  $\delta \approx 5 \cdot 10^{-3}$.
Calculating with the BS effective distribution of the present
paper, we find $\delta \approx 0.7 \cdot 10^{-3}$. Note, that  in
ref.~\cite{urel} the quantity $\delta \approx 1 \cdot 10^{-2}$
has been reported. We attribute this small discrepancy to the
less sophisticated numerical approximation which had been made
in~\cite{urel}. This approximation slightly underestimates a high
momentum behavior of $f^{N/D}$.

\subsection{Polarized deuteron, $b_1^D$ and $b_2^D$}

The SFs of the deuteron $b_{1,2}^D(x)$ are calculated within both
the relativistic   and nonrelativistic approaches.  The
relativistic calculations are based on the formulae (\ref{fndm}),
(\ref{fb1})-(\ref{dfnd}). The  nonrelativistic calculations,
eq.~(\ref{dfndnr}), uses the realistic wave function of the
deuteron in the Bonn potential~\cite{bonn}. Another ingredient of
the calculations, the nucleon  SFs  $F_{1,2}^N(x,Q^2)$, is again
taken from ref.~\cite{amb} at $Q^2 = 10$~GeV$^2$. The results are
neither very sensitive to the particular choice of the
parametrization for  the nucleon  SFs nor to the $Q^2$-dependence
of them.

The distribution functions $\Delta f^{N/D} (y)$  are calculated
and the results are presented in Fig.~\ref{dfndp}. The
relativistic (solid line) and nonrelativistic (dotted line)
calculations give  similar behavior of the distribution function.
Indeed, it is difficult to distinguish between them, not speaking
about making  definite conclusions.  The third line in the
Fig.~\ref{dfndp} is given for  illustration, and presents
$y\Delta f^{N/D}(y)$  for the relativistic calculations. The
calculation of the sum rules is more representative. To
understand the scale of effects, which are discussed below,  it
is customary to define auxiliary quantities:  \begin{eqnarray}
\int\limits_{0}^{M_D/m} {\sf Abs}\left ( \Delta f^{N/D}(y)\right
)dy  \simeq  \int\limits_{0}^{M_D/m} {\sf Abs}\left ( y \Delta
f^{N/D}(y)\right )dy \simeq 0.14. \label{aux} \end{eqnarray} The
Bethe-Salpeter   and   nonrelativistic Bonn calculations give the
same result  in (\ref{aux}), with accuracy of $\sim 5\%$. Thus,
the effective distribution functions, $ \Delta f^{N/D}$, are an
order  of magnitude smaller than the usual spin-independent
distributions $f^{N/D}$ normalized to 1. This is not a very
important circumstance, but it works against accuracy in the
numerical calculations, since $ \Delta f^{N/D}$ is a difference
of two functions normalized to 1 ($M=1$ and $M=0$).  Numerically,
the   sum rule (\ref{sr1}) (see also eq.~(\ref{sr1nr})) is
satisfied both in relativistic and nonrelativistic calculations
with good accuracy, despite the approximate numerical ``inverse
Wick rotation'' and the discussion after eq.~(\ref{dfndnr}).  The
corresponding integrals are $\sim 5\cdot 10^{-4}$   and $\sim
3\cdot10^{-5}$ and they should be compared  to the
estimate~(\ref{aux}). The sum rule  (\ref{ch0})  may be used to
improve distributions  $\Delta f^{N/D}$ by making integrals for
$f^{N/D}_1$ and  $f^{N/D}_0$ {\em exactly } the same. However,
this does not lead to a significant variation of results for
SFs, except $x\to 0$ for $b_1^D(x)$.

The behavior of  $b_1^D(x)$ at $x\to 0$ deserves to be considered
more closely, especially for  numerical calculations, since the
nucleon function $F_1^N(x)$  can be divergent at small $x$.
Unfortunately it is impossible to estimate $b_1^D(0)$ exactly for
the realistic  SF $F_1^N$. However, a contribution of the
singularity can be evaluated. Indeed, let us assume a singular
behavior as $F_1^N\sim C/x$, then eq.~(\ref{fb1}) leads for small
$x$ to \begin{eqnarray} \!\!\!\!\!\!\!\! b_1^D (x\to 0)  &\sim&
\frac{C}{x} \int\limits_{x}^{M_D/m} \Delta f^{N/D} (y)dy  =
\frac{C}{x}  \left\{ \int\limits_{0}^{M_D/m}  -
\int\limits_{0}^{x} \right \} \Delta f^{N/D} (y) dy\nonumber \\
&\sim& \frac{C\cdot Z}{x} - C \Delta f^{N/D} (0), \label{estd}
\end{eqnarray} where $Z = 0$ in exact relativistic formula, but
it can be a small number in numerical  calculations or in the
nonrelativistic formalism.  Thus, the limit of the deuteron  SF
$b_1^D(x)$ as $x \to 0$  is a constant, but one has to exercise
great care in performing  numerical computations, since any error
leads to a divergent behavior  at small $x$. In this context,  an
adjustment of norms of the two terms in formulae (\ref{dfnd}) and
(\ref{dfndnr}) has a meaning of subtraction of  the numerical
error from $b_1^D$ at small $x$.

The situation with the second sum rule (\ref{sr2}) is quite
different. Numerically it is violated more significantly than the
previous one. Corresponding integrals are $\sim 1\cdot 10^{-3}$
and $\sim 3\cdot 10^{-3}$  for relativistic and nonrelativistic
calculations respectively, i.e. about $0.7\%$ and $2\%$ compare
to (\ref{aux}). Therefore,  numerical approximations slightly
damage the relativistic formula too. It is attributed to the
numerical rotation in the Minkowski space. An adjustment of the
normalization, as it has been discussed, slightly improves  the
accuracy (to $0.5\%$). On the contrary, the result for the
nonrelativistic approach is stable with respect to any
adjustments, since it is defined by the formulae (\ref{sr2nr}).

The  SFs $b_1^D$ and $b_2^D$ are calculated within  two
approaches as well. The results are shown in Fig.~\ref{b2p} a)
and b). The behavior of the functions in Fig.~\ref{b2p} a)
suggests the validity of the  sum rule (\ref{sr1}). At the same
time,  the nonrelativistic calculation for  $b_2^D$ in
Fig.~\ref{b2p} b) (dotted line) obviously does not satisfy the
sum rule (\ref{sr2}). The main difference of the relativistic and
nonrelativistic calculations is at small $x$, where these
approaches give different signs for the  SFs. To illustrate the
effect of the presence of the $\Theta$-function under integral
in  the nonrelativistic formula (\ref{dfndnr}), the calculations
have been done as well with a restricted interval of integration
over  $|{\bf p}|$. The condition $|{\bf p}|< 0.7 $~GeV
corresponds to the ``softer'' deuteron wave function, but makes
the sum rule (\ref{sr2nr}) exact.  Corresponding  SFs are shown
in Fig.~\ref{b2p} a) and b) (dashed line). The result of this
``experiment'' is that the effect of $\Theta$-function is not
quantitatively significant.  It also does not affect the
principle conclusion about the second sum rule (\ref{sr2}), but
makes the defect a tiny bit smaller. This is understandable,
since the sum rule breaking terms in (\ref{sr2nr})  is $\propto
|{\bf p}|cos\theta$.

\subsection{Polarized deuteron, $g_1^D$}

The spin-dependent SF of the deuteron, $g_1^D$, is calculated
using the same three models as the spin-independent SF, $F_2$:
the fully relativistic BS approach (solid line), nonrelativistic
approach based on the Bonn wave function(dotted line) and
nonrelativistic approach which uses the exact density of the BS
approach (dashed line).  The nucleon SF, $g_1^N$ is taken from
ref.~\cite{shaf}. The results of calculations are presented in
Fig.~\ref{g1}.

For  illustration we also present in  Fig.~\ref{g1} the quantity
$\langle \gamma_5\gamma_3\rangle^{BS}_{M=1}$ which corresponds
to a ``model'' for the deuteron SF (dot-dashed straight line ):
\begin{eqnarray} g_1^D(x,Q^2) = \frac{1}{4\pi}\langle
\gamma_5\gamma_3\rangle^{BS}_{M=1}\cdot g_1^N(x,Q^2).
\label{expmod} \end{eqnarray} If we replace $\langle
\gamma_5\gamma_3\rangle^{BS}_{M=1}$  by the factor $\left
(1-3/2\cdot w_D\right )$ from eq.~(\ref{naxnr}), we get the
formula usually used by experimentalists  aiming to obtain the
neutron SF, $g_1^n$ from the combined proton and deuteron data.

Fig.~\ref{g1} shows that picture is different from the naive
estimate (\ref{expmod}). However, within the present day
experimental errors it may be {\em a reasonable approximation}
(see e.g. ref.~\cite{urel2}). Huge  leaps of the ratio around
the constant  $\langle \gamma_5\gamma_3\rangle^{BS}_{M=1}$ at
$x<0.7$ are not too important. They correspond to  zeros of the
nucleon SF which are slightly shifted by the  convolution
formula. Systematic difference in the ratio exist between the
nonrelativistic calculation (dotted line) and two calculations
based on the BS  densities (solid and dashed curves). This is a
difference between $D$-wave admixture in the Bonn potential ($w_D
\approx 4.3\%$) and our solution of the BS equation ($w_D \approx
5\%$).  The rise of the ratio,  at $x$ higher than 0.7, is of the
same nature as in the spin-independent case in Fig.~\ref{rf2}, it
is caused by the Fermi motion.

\section{Summary} \label{sec:sum}

In this paper, we have  presented a  study of  the deep inelastic
electron scattering on the deuteron in the Bethe-Salpeter
formalism in the realistic  meson-nucleon model. In particular,
\begin{itemize}

\item The connection of the structure functions to the densities
of the appropriate charges and currents is analyzed. By analysing
the same densities in the nonrelativistic approach, we have
systematically compared   the relativistic and  nonrelativistic
calculations, and established sources of the relativistic effects.

\item  Using our numerical solution of the Bethe-Salpeter
equation amplitude with a realistic kernel, the leading twist
structure functions of the deuteron, $F_{2}^D$, $b_{1,2}^D$ and
$g_1^D$, are  calculated in   the fully relativistic fashion.

\item  Our numerical calculations of the structure functions
emphasize a qualitative agreement with previous non-relativistic
results.  However, we have found  effects systematically
distinguishing  a consistent relativistic approach from the
nonrelativistic one: in the  relativistic formalism (i) the
magnitude of binding effects is larger and (ii) the effect of
Fermi motion at high $x$ is stronger; and (iii) the
nonrelativistic calculations suffer   unavoidable internal
inconsistencies, which lead to small effects in the structure
functions $F_2^D$ and $g_1^D$, but seriously damage structure
function $b_{1,2}^D$  and  noticably break sum rules for this
function. \end{itemize}

The present paper concludes our systematic study of the deep
inelastic electron (muon) scattering on the deuteron  in the
Bethe-Salpeter formalism. The results are collected in the two
papers, ref.~\cite{1}  and the present paper, and in part have
been also published previously in
refs.~\cite{urel,urel1,urel2,kuco}. The main lesson, we have
learned from this study, is that the deuteron in the deep
inelastic reaction indeed behaves as a very  slightly
relativistic system. One has to look for special  conditions or
kinematics of the reaction to be able to find  noticible
relativistic effects.  We have found  certain situations where
the relativistic approach is absolutely essential and use  of the
nonrelativistic methods is not justified. Most representative
examples are the high $x$ behaviour of the structure funtions and
the spin-dependent structure functions, $b^D_{1,2}$.

Apart from the phenomenological differences between the
relativistic and nonrelativistic approaches, the most important
merit of the covariant formalism is its {\em consistency}. Its
close connection to the field theory guaranties that the
calculated observables obey  the  sum rules and other general
properties imposed by the fundamental principles. In this sense
the relativistic approach is definitely more advanced
theoretically than its nonrelativistic counterpart and  provides
a better understanding of the deep inelastic scattering on the
deuteron.


\section*{Acknowledgments}


We would like to thank F. Gross, W. Van Orden and J. Tjon for
conversations  about properties of the relativistic equations and
its solutions. Authors also wish to thank  everybody who had
discussed various topics of this paper with us. In particular, we
would like to thank  S. Bondarenko, V. Burov, C. Ciofi degli
Atti, S. Dorkin,  A. Efremov, B. K\"ampfer, K. Kazakov, S.
Kulagin, W. Melnitchouk,   A. Molochkov, S. Scopetta,  C. Shakin,
S. Simula, O. Teryaev,  A. Titov, A. Thomas and W. Weise.

The research was supported in part by the Natural Sciences and
Engineering Research Council of Canada and by Instituto Nazionale
di Fisica Nucleare, Sezione di Perugia, Italy.


\appendix

\section{Explicit formulae for densities} \label{sec:dens}

In this Appendix we present formulae which allow to restore
explicit form of the various densities in the BS formalism. For
convenience, we define auxiliary ``densities'': \begin{eqnarray}
&& \left \{ \begin{array}{c} f_0^{N/D}(p) \\ f_3^{N/D}(p)
\end{array} \right \} = \frac{ 1}{3}{\sum_M}\int\limits_0^{2\pi}
{d\phi} {\sf Tr}\left\{ \bar\Psi_M(p_0,{\bf  p})\right. \left \{
\begin{array}{c} \gamma_0 \\ \gamma_3  \end{array} \right \}
\left. \phantom{\bar\Psi}\!\!\!\!\!\! \Psi_M(p_0,{\bf  p}) (\hat
p_2-m) \right \}, \label{auxun}\\ && \left \{ \begin{array}{c}
\Delta f_0^{N/D}(p) \\ \Delta f_3^{N/D}(p) \end{array} \right \}
= \int\limits_0^{2\pi} {d\phi} {\sf Tr}\left\{
\bar\Psi_{M}(p_0,{\bf  p})\right. \left \{ \begin{array}{c}
\gamma_0 \\ \gamma_3  \end{array} \right \} \left.
\phantom{\bar\Psi}\!\!\!\!\!\! \Psi_{M}(p_0,{\bf  p}) (\hat
p_2-m) \right \}_{M=1} -  \int\limits_0^{2\pi} {d\phi} {\sf
Tr}\left\{ \ldots \right \}_{M=0} , \label{auxb}\\ && \left \{
\begin{array}{c} \vec f_0^{N/D}(p) \\ \vec f_3^{N/D}(p)
\end{array} \right \} = \int\limits_0^{2\pi} {d\phi} {\sf
Tr}\left\{ \bar\Psi_{M}(p_0,{\bf  p})\right. \left \{
\begin{array}{c} \gamma_5\gamma_0 \\ \gamma_5\gamma_3
\end{array} \right \} \left. \phantom{\bar\Psi}\!\!\!\!\!\!
\Psi_{M}(p_0,{\bf  p}) (\hat p_2-m) \right \}_{M=1},
\label{auxpo} \end{eqnarray} where integration over $\phi$ leads
to the trivial factor of $2\pi$, since none of the matrix
elements on the r.h.s.  depends on $\phi$. To obtain explicit
expressions for the densities discussed  in the paper we have to
compare  eq.~(\ref{auxun})-(\ref{auxpo}) with definitions of
corresponding densities.

Note, that a formalism  presented in refs.~\cite{1,urel} and in
this paper can be   easily adopted for  an analytical computer
calculations.  The following  results have been obtained
utilizing the {\em Mathematica} package~\cite{wolf}:

{   \begin{eqnarray}  f_{0}^{N/D}(p) = &m&\left( -8
\psi_{a0}(p_0,p) \psi_{t0}(p_0,p) -  8 \psi_{a2}(p_0,p)
\psi_{t2}(p_0,p) \right) \label{a1} \\  &+\quad \quad
p\quad\quad\quad&  \left( {{-8 \psi_{p1}(p_0,p)
\psi_{t0}(p_0,p)}\over  {{\sqrt{3}}}} + 8 {\sqrt{{2\over 3}}}
\psi_{p1}(p_0,p)  \psi_{t2}(p_0,p) \right. \nonumber\\ &&
\quad\left. +  4 {\sqrt{{2\over 3}}} \psi_{a0}(p_0,p)
\psi_{v1}(p_0,p) +  {{4 \psi_{a2}(p_0,p) \psi_{v1}(p_0,p)}\over
{{\sqrt{3}}}}\ \right)\nonumber  \\ &+ \left( {{1}\over 2}M_d -
p_0 \right)  & \left( 2 {{\psi_{a1}^0(p_0,p)}^2} + 2
{{\psi_{a0}(p_0,p)}^2} +  2 {{\psi_{a2}(p_0,p)}^2} + 2
{{\psi_{p1}(p_0,p)}^2} \right.\nonumber \\ &&\quad\left. +  8
{{\psi_{t0}(p_0,p)}^2} + 8 {{\psi_{t2}(p_0,p)}^2} +  8
{{\psi_{t1}^0(p_0,p)}^2} + 2 {{\psi_{v1}(p_0,p)}^2} \right)
\nonumber \end{eqnarray} \begin{eqnarray}  f_{3}^{N/D}(p) =
&&\cos (\theta) \left\{ \phantom{{M_d}\over 2}\right.
\label{a2}\\ & \left( {{1}\over 2}M_d - p_0 \right)&  \left( {{8
\psi_{p1}(p_0,p) \psi_{t0}(p_0,p)}\over  {{\sqrt{3}}}} - 8
{\sqrt{{2\over 3}}} \psi_{p1}(p_0,p)  \psi_{t2}(p_0,p)
\right.\nonumber\\  &&\quad \left. - 4 {\sqrt{{2\over 3}}}
\psi_{a0}(p_0,p)  \psi_{v1}(p_0,p) -  {{4 \psi_{a2}(p_0,p)
\psi_{v1}(p_0,p)}\over  {{\sqrt{3}}}} \right)  \nonumber \\
&+\quad\quad m\quad\quad\quad&   \left( {{-4 \psi_{a0}(p_0,p)
\psi_{p1}(p_0,p)}\over {{\sqrt{3}}}} +  4 {\sqrt{{2\over 3}}}
\psi_{a2}(p_0,p)  \psi_{p1}(p_0,p)  \right . \nonumber\\
&&\quad \left.+  8 {\sqrt{{2\over 3}}} \psi_{t0}(p_0,p)
\psi_{v1}(p_0,p) +  {{8 \psi_{t2}(p_0,p) \psi_{v1}(p_0,p)}\over
{{\sqrt{3}}}} \right) \nonumber \\  &+ \quad\quad
p\quad\quad\quad&   \left( 2 {{\psi_{a1}^0(p_0,p)}^2} -  {{2
{{\psi_{a0}(p_0,p)}^2}}\over 3} -  {{8 {\sqrt{2}}
\psi_{a0}(p_0,p) \psi_{a2}(p_0,p)}\over 3} \right.\nonumber \\
&&\quad \left. + {{2 {{\psi_{a2}(p_0,p)}^2}}\over 3}  -  2
{{\psi_{p1}(p_0,p)}^2} +  {{8 {{\psi_{t0}(p_0,p)}^2}}\over 3} - 2
{{\psi_{v1}(p_0,p)}^2} \right.\nonumber\\  &&\quad \left.  +
{{32 {\sqrt{2}} \psi_{t0}(p_0,p)  \psi_{t2}(p_0,p)}\over 3} -
{{8 {{\psi_{t2}(p_0,p)}^2}}\over 3}  +  8
{{\psi_{t1}^0(p_0,p)}^2} \right) \left. \! \! \! \! \! \! \! \!
\! \! \phantom{{M_d}\over 2}\right \} \nonumber \end{eqnarray}}

{   \begin{eqnarray}  \vec f_{0}^{N/D}(p) = &&\cos (\theta)
\left\{\phantom{{{M_d}\over 2}} \right.\label{a3} \\ &\left(
{{1}\over 2}M_d - p_0 \right)  &\!\!\!\!\!\! \left( 2 {\sqrt{6}}
\psi_{a0}(p_0,p) \psi_{v1}(p_0,p) +  2 {\sqrt{3}}
\psi_{a2}(p_0,p) \psi_{v1}(p_0,p) \right) \nonumber \\[2mm]
&+\quad\quad m\quad\quad\quad &\!\!\!\!\!\!\left( -4 {\sqrt{6}}
\psi_{t0}(p_0,p)  \psi_{v1}(p_0,p) -  4 {\sqrt{3}}
\psi_{t2}(p_0,p) \psi_{v1}(p_0,p) \right)\nonumber\\[2mm]
&+\quad\quad  p\quad\quad\quad &\!\!\!\!\!\! \left( 2
{{\psi_{a0}(p_0,p)}^2} +  2 {\sqrt{2}} \psi_{a0}(p_0,p)
\psi_{a2}(p_0,p) +  {{\psi_{a2}(p_0,p)}^2} \right.\nonumber \\ &&
\left .  - 8 {{\psi_{t0}(p_0,p)}^2} -  8 {\sqrt{2}}
\psi_{t0}(p_0,p) \psi_{t2}(p_0,p) -  4 {{\psi_{t2}(p_0,p)}^2}
\right.\nonumber \\ && \left . - 12 {{\psi_{t1}^0(p_0,p)}^2} +  3
{{\psi_{v1}(p_0,p)}^2} \right)  \left.\phantom{{{M_d}\over 2}}
\hspace*{-.5cm}\right\}\nonumber \end{eqnarray} \begin{eqnarray}
\vec f_{3}^{N/D}(p) = &&P_2(\cos (\theta))  \left\{
\phantom{\frac{1^1_1}{1^1_1}} \right. \label{a4} \\
&\quad\quad\quad p\quad\quad\quad&  \left( {{-8 \psi_{p1}(p_0,p)
\psi_{t0}(p_0,p)}\over {{\sqrt{3}}}} -  4 {\sqrt{{2\over 3}}}
\psi_{p1}(p_0,p)  \psi_{t2}(p_0,p)  \right. \nonumber\\ && \left.
\quad -  2 {\sqrt{{2\over 3}}} \psi_{a0}(p_0,p)  \psi_{v1}(p_0,p)
-  {{8 \psi_{a2}(p_0,p) \psi_{v1}(p_0,p)}\over  {{\sqrt{3}}}}
\right)\nonumber \\[2mm] &+\quad\quad m\quad\quad\quad& \left( 4
{\sqrt{2}} \psi_{a2}(p_0,p)  \psi_{t0}(p_0,p) +  4 {\sqrt{2}}
\psi_{a0}(p_0,p) \psi_{t2}(p_0,p) \right. \nonumber\\ && \left.
\quad +  8 \psi_{a2}(p_0,p) \psi_{t2}(p_0,p)   +  4 {\sqrt{2}}
\psi_{a1}^0(p_0,p) \psi_{t1}^0(p_0,p) \right.\nonumber \\  &&
\left. \quad  -  2 {\sqrt{2}} \psi_{p1}(p_0,p) \psi_{v1}(p_0,p)
\right)\nonumber \\[2mm]  & +  \left( {{1}\over 2} M_d- p_0
\right)  &  \left( -2 {\sqrt{2}} \psi_{a0}(p_0,p)
\psi_{a2}(p_0,p) -  2 {{\psi_{a2}(p_0,p)}^2} \right. \nonumber\\
&& \left. \quad-  8 {\sqrt{2}} \psi_{t0}(p_0,p) \psi_{t2}(p_0,p)
-  8 {{\psi_{t2}(p_0,p)}^2} \right. \nonumber\\ && \left. \quad -
8 {{\psi_{t1}^0(p_0,p)}^2} -  2 {{\psi_{v1}(p_0,p)}^2} \right)
\hspace*{-.3cm} \left.\phantom{\frac{1^1_1}{1^1_1}} \right\}
\nonumber \\[4mm] &+\quad\quad p\quad\quad\quad& \left( {{8
\psi_{p1}(p_0,p) \psi_{t0}(p_0,p)}\over  {{\sqrt{3}}}} + 4
{\sqrt{{2\over 3}}} \psi_{p1}(p_0,p)
\psi_{t2}(p_0,p)\right.\nonumber \\ && \left. \quad -  4
{\sqrt{{2\over 3}}} \psi_{a0}(p_0,p)  \psi_{v1}(p_0,p) +  {{2
\psi_{a2}(p_0,p) \psi_{v1}(p_0,p)}\over {{\sqrt{3}}}}\ \right)
\nonumber  \\[2mm] &+\quad\quad m\quad\quad\quad& \left( 8
\psi_{a0}(p_0,p)  \psi_{t0}(p_0,p) -  4 \psi_{a2}(p_0,p)
\psi_{t2}(p_0,p) \right.\nonumber \\ && \left. \quad -  4
{\sqrt{2}} \psi_{a1}^0(p_0,p) \psi_{t1}^0(p_0,p) +  2 {\sqrt{2}}
\psi_{p1}(p_0,p) \psi_{v1}(p_0,p) \right)\nonumber \\[2mm] & +
\left( {{1}\over 2} M_d- p_0 \right)  & \left( -2
{{\psi_{a0}(p_0,p)}^2} + {{\psi_{a2}(p_0,p)}^2} -  8
{{\psi_{t0}(p_0,p)}^2} \right.\nonumber \\ && \left. \quad  + 4
{{\psi_{t2}(p_0,p)}^2} -  4 {{\psi_{t1}^0(p_0,p)}^2} -
{{\psi_{v1}(p_0,p)}^2} \right)\nonumber \end{eqnarray}}

{   \begin{eqnarray}  \Delta f_{0}^{N/D}(p) = && P_2(\cos
(\theta)) \left\{\phantom{\frac{1^1_1}{1^1_1}}  \right.\label{a5}
\\ &\quad\quad\quad m\quad\quad\quad& \left( -12 {\sqrt{2}}
\psi_{a2}(p_0,p) \psi_{t0}(p_0,p) -  12 {\sqrt{2}}
\psi_{a0}(p_0,p) \psi_{t2}(p_0,p) \right. \nonumber \\ && \left.
\quad+  12 \psi_{a2}(p_0,p) \psi_{t2}(p_0,p)
{\phantom{\sqrt{2}}}\!\!\!\!\!\!\!\!\right) \nonumber\\
&+\quad\quad p\quad\quad\quad& \left( 8 {\sqrt{3}}
\psi_{p1}(p_0,p)  \psi_{t0}(p_0,p) -  8 {\sqrt{6}}
\psi_{p1}(p_0,p) \psi_{t2}(p_0,p) \right.\nonumber \\ && \left.
\quad+  2 {\sqrt{6}} \psi_{a0}(p_0,p) \psi_{v1}(p_0,p) +  2
{\sqrt{3}} \psi_{a2}(p_0,p) \psi_{v1}(p_0,p) \right) \nonumber\\
& +  \left( {{1}\over 2} M_d- p_0 \right)  &   \left( -6
{{\psi_{a1}^0(p_0,p)}^2} +  6 {\sqrt{2}} \psi_{a0}(p_0,p)
\psi_{a2}(p_0,p) -  3 {{\psi_{a2}(p_0,p)}^2} \right.\nonumber \\
&& \left. \quad - 6 {{\psi_{p1}(p_0,p)}^2} +  24 {\sqrt{2}}
\psi_{t0}(p_0,p) \psi_{t2}(p_0,p) -  12 {{\psi_{t2}(p_0,p)}^2}
\right.\nonumber \\ && \left. \quad +  12
{{\psi_{t1}^0(p_0,p)}^2} + 3 {{\psi_{v1}(p_0,p)}^2} \right)
\hspace*{-4mm}\left. \phantom{\frac{1^1_1}{1^1_1}} \right\}
\nonumber \end{eqnarray} \begin{eqnarray}  \Delta f_{3}^{N/D}(p)
= &&\cos (\theta) \left\{\phantom{\frac{1^1_1}{1^1_1}}
\right.\label{a6}\\ &\quad\quad\quad p\quad\quad\quad& \left( -4
{{\psi_{a0}(p_0,p)}^2} +  2 {\sqrt{2}} \psi_{a0}(p_0,p)
\psi_{a2}(p_0,p) +  4 {{\psi_{a2}(p_0,p)}^2} \right. \nonumber\\
&& \left. \quad + 16 {{\psi_{t0}(p_0,p)}^2} -  8 {\sqrt{2}}
\psi_{t0}(p_0,p) \psi_{t2}(p_0,p) -  16 {{\psi_{t2}(p_0,p)}^2}
\right)\nonumber \\[2mm] & +  \left( {{1}\over 2} M_d- p_0
\right)  &   \left( -8 {\sqrt{3}} \psi_{p1}(p_0,p)
\psi_{t0}(p_0,p) -  4 {\sqrt{6}} \psi_{p1}(p_0,p)
\psi_{t2}(p_0,p) \right.\nonumber \\ && \left. \quad -  2
{\sqrt{6}} \psi_{a0}(p_0,p) \psi_{v1}(p_0,p) +  4 {\sqrt{3}}
\psi_{a2}(p_0,p) \psi_{v1}(p_0,p) \right)\nonumber \\[2mm]
&+\!\!\!\quad\quad m\quad\quad\quad& \left( 4 {\sqrt{3}}
\psi_{a0}(p_0,p)  \psi_{p1}(p_0,p) +  2 {\sqrt{6}}
\psi_{a2}(p_0,p) \psi_{p1}(p_0,p) \right. \nonumber\\ && \left.
\quad+  4 {\sqrt{6}} \psi_{t0}(p_0,p) \psi_{v1}(p_0,p) -  8
{\sqrt{3}} \psi_{t2}(p_0,p) \psi_{v1}(p_0,p) \right)
\nonumber\\[4mm] &&\hspace*{-3.1cm} + P_2(\cos (\theta))
\left[\phantom{\frac{1^1_1}{1^1_1}}\right.\nonumber \\ &+  \left(
{{1}\over 2} M_d- p_0 \right)& \left( 12 {\sqrt{6}}
\psi_{p1}(p_0,p)  \psi_{t2}(p_0,p) -  6 {\sqrt{3}}
\psi_{a2}(p_0,p) \psi_{v1}(p_0,p) \right)\nonumber\\[2mm]
&+\!\!\!\quad\quad m\quad\quad\quad& \left( -6 {\sqrt{6}}
\psi_{a2}(p_0,p)  \psi_{p1}(p_0,p) +  12 {\sqrt{3}}
\psi_{t2}(p_0,p) \psi_{v1}(p_0,p) \right)\nonumber\\[2mm]
&+\quad\quad p\quad\quad\quad& \left( -6 {{\psi_{a1}^0(p_0,p)}^2}
-  9 {{\psi_{a2}(p_0,p)}^2} + 6 {{\psi_{p1}(p_0,p)}^2}
\right.\nonumber \\ && \left. \quad+  36 {{\psi_{t2}(p_0,p)}^2}
+  12 {{\psi_{t1}^0(p_0,p)}^2} - 3 {{\psi_{v1}(p_0,p)}^2}\
\right) \hspace*{-.3cm}
\left.\phantom{\frac{1^1_1}{1^1_1}}\right]   \hspace*{-.6cm}
\left. \phantom{\left.\frac{1^1_1}{1^1_1}\right]} \right\}
\nonumber \end{eqnarray}}

\section{Hadronic tensor for the deuteron. Projectors.}
\label{sec:hadron}

The parametrization of   hadron tensor for the deuteron, utilized
in the present paper, is given by eq.~(\ref{htend}). It has both
symmetric, $\{\ldots\}$, and antisymmetric  $[\ldots]$ parts in
respect to the permutation of its indices: \begin{eqnarray}
W_{\mu\nu}^D = W_{\{\mu\nu\}}^D + W_{[\mu\nu]}^D.\label{ht2}
\end{eqnarray}

Three physical vectors are used in this parametrization:
\begin{enumerate} \item $P_D$ the deuteron momentum. In the rest
frame of the deuteron $P_D = (M_D,{\bf 0})$; \item $q$ the
momentum transfer in the deep inelastic scattering.  With proper
choice of the orientation of the coordinate system
$q=(\nu,0,0,-\sqrt{\nu^2+Q^2})$. In the deep inelastic limit,
when $Q^2/\nu^2 \to 0$, $pq = \nu(p_0+p_3)$. \item $S_D(M)$ is
the total angular momentum of the deuteron, i.e. spin of the
deuteron as an elementary particle: \begin{eqnarray}
S_{D}^\alpha(M) =
-\frac{i}{M_D}\epsilon^{\alpha\beta\gamma\delta}
E^*_\beta(M)E_\gamma(M)P_{D\delta}, \label{spin1}\\[3mm] E(M) =
\left \{ \begin{array}{cl} \frac{1}{\sqrt{2}}(0,-1,-i,0), & \quad
M = 1\\ (0,0,0,1),  & \quad M = 0\\
\frac{1}{\sqrt{2}}(0,1,-i,0),  & \quad M = -1 \end{array}
\right.\label{ee} \end{eqnarray} \end{enumerate}

The symmetric part of the hadron tensor, $W_{\{\mu\nu\}}^D$,
contains terms proportional to the two following tensor
structures: \begin{eqnarray} && T^{(1)}_{\mu\nu} = -g_{\mu\nu} +
\frac{q_\mu q_\nu}{q^2},\label{t1}\\ && T^{(2)}_{\mu\nu} =  \left
( P_{D\mu} - q_\mu \frac{P_Dq}{q^2}  \right )  \left ( P_{D\nu} -
q_\nu \frac{P_Dq}{q^2}  \right ) \frac{1}{P_Dq}. \label{t2}
\end{eqnarray} Because of the conservation of the electromagnetic
current, the contraction of  the hadron tensor with  $q_{\mu}$ in
respect of any index is zero. That is why  only
$P_{D\mu}P_{D\nu}$ and $g_{\mu\nu}$ are available to construct
the  projection operators to extract structure functions $F_1^D$
and $F_2^D$ from the  hadron tensor. We introduce the following
coefficients: \begin{eqnarray} && C_1 \equiv
g^{\mu\nu}T^{(1)}_{\mu\nu}, \quad C_2 \equiv
g^{\mu\nu}T^{(2)}_{\mu\nu}, \quad C_3 \equiv \frac{P_{D}^\mu
P_{D}^\nu}{P_D^2} T^{(1)}_{\mu\nu},  \quad C_4 \equiv
\frac{P_{D}^\mu P_{D}^\nu}{P_D^2} T^{(2)}_{\mu\nu},
\label{cccc}\\[3mm] && D_C \equiv C_1\cdot C_4 - C_2 \cdot C_3,
\label{dc} \\[3mm] && A_1 \equiv g^{\mu\nu} W_{\mu\nu}^D, \quad
A_2 \equiv  \frac{P_{D}^\mu P_{D}^\nu}{P_D^2} W_{\mu\nu}^D
\label{aa}. \end{eqnarray} Then the structure function are
recovered by: \begin{eqnarray} F_1^D = \frac{A_1\cdot C_4 - A_2
\cdot C_2}{D_C}, \label{exf1}\\ F_2^D = \frac{A_2\cdot C_1 - A_1
\cdot C_3}{D_C}. \label{exf2} \end{eqnarray}

The antisymmetric part of the hadron tensor of the deuteron,
$W_{[\mu\nu]}^D$, in the general case has a form:
\begin{eqnarray} W_{[\mu\nu]}^D = && \frac{iM_D}{P_Dq}
\epsilon_{\mu\nu\alpha\beta}  q^\alpha \left \{ S_D^\beta(M)
\left (g_1^D(x_D,Q^2) + g_2^D(x_D,Q^2) \right ) -P_D^\beta
\frac{(S_D(M)q)}{P_Dq}g_2^D(x_D,Q^2) \right \} , \label{want}
\end{eqnarray} where the second structure function, $g_2^D$, is
vanishing in the  deep inelastic limit, $ \nu \to \infty, \quad
Q^2 \to \infty ,\quad Q^2/\nu \to const$. We do not discuss this
structure function in the present paper. To obtain the
spin-dependent  structure function, $g_1^D$, we construct the
antisymmetric projectors. The two following projectors  are
equivalent for our purpose: \begin{eqnarray} && R^{(1)}_{\mu\nu}
\equiv i\epsilon_{\mu\nu\alpha\beta}q^\alpha S^\beta_D(M),
\label{aw1}\\ && R^{(2)}_{\mu\nu} \equiv \frac{i (S_D(M)q)}{P_Dq}
\epsilon_{\mu\nu\alpha\beta}q^\alpha P_D^\beta . \label{aw2}
\end{eqnarray} It is interesting that  in the limit $Q^2/\nu^2
\to 0$: \begin{eqnarray} g_1^D = \frac{R^{(1)\mu\nu}W_{\mu\nu}^D
}{2\nu} = \frac{R^{(2)\mu\nu}W_{\mu\nu}^D }{2\nu} . \label{exg1}
\end{eqnarray}


\newpage

\begin{center}  \begin{tabular}{|c|c|c|c|c|}  \hline meson &
coupling constants       & mass     & cut-off & isospin   \\ B
& $g^2_B/(4\pi); [g_v/g_t]$ &$\mu_B$, GeV  &$\Lambda$,GeV &
\\ \hline \hline $\sigma$ & 12.2 & 0.571 & 1.29 & 0  \\ \hline
$\delta$ & 1.6 & 0.961 & 1.29  & 1 \\ \hline $\pi$ & 14.5 & 0.139
& 1.29   & 1\\ \hline $\eta$ & 4.5 & 0.549 & 1.29  & 0 \\ \hline
$\omega$ & 27.0; [0] & 0.783 & 1.29  & 0 \\ \hline $\rho$ &  1.0;
[6] & 0.764 & 1.29 & 1  \\ \hline \hline \multicolumn{5}{|c|}{$m
= 0.939$ GeV, $\varepsilon_D = -2.225$ MeV}\\ \hline
\end{tabular} \end{center}

\vskip 5mm

{\bf Table 1.} {\em The set of parameters of the kernel of BS
equation used in this work.}

\newpage


\begin{figure} \caption{Triangle graph for nucleon contribution
to the deuteron matrix element of operator $\hat {\cal O}$.}
\label{tre-dia} \end{figure}

\begin{figure} \caption{The charge density in the deuteron,
eqs.~(\ref{nchbs}) and (\ref{nchnr}) calculated in different
models, nonrelativistic and relativistic.  Curves: the
Bethe-Salpeter amplitude  (solid), the Bonn wave function
(dashed), the Paris wave function   (dotted).} \label{g0den}
\end{figure}

\begin{figure} \caption{The $\langle \gamma_3  \rangle ({\bf
p})$  in the deuteron calculated  in different models. Curves:
BS  amplitude  (solid), the Bonn wave function  (dashed), the
Paris wave function  (dotted).} \label{g3den} \end{figure}

\begin{figure} \caption{The ``tensor density'' of the deuteron,
$\langle \gamma_0 \rangle^{NR}_{M=1}({\bf p}) -\langle \gamma_0
\rangle^{NR}_{M=0}({\bf p})$, calculated  in different models. To
exclude the angular dependence the densities are divided by
$P_2(cos\theta)$. Curves: BS  amplitude  (solid), the Bonn wave
function (dashed), the Paris wave function  (dotted).}
\label{dg0den} \end{figure}

\begin{figure} \caption{The spin density, $\langle
\gamma_5\gamma_3\rangle_{M=1} ({\bf p})$,  in the deuteron
calculated  in different models (see Section~\ref{sec:statden}
for definitions).  To exclude the angular dependence for the
present figure,  densities are integrated over all angles.
Curves: BS  amplitude  (solid), the Bonn wave function  (dashed),
the Paris wave function  (dotted).} \label{g53den} \end{figure}


\begin{figure} \caption{The one-pole contribution  to the charge
density  of the deuteron, $\langle \gamma_0\rangle
^{BS}_{pole}({\bf p})$, calculated with the BS amplitude. Curves:
the leading term in the Taylor expansion (dotted), first two
terms, up to $\propto p_2^2$ (dashed), first three terms, up to
$\propto p_4^2$ (solid).  } \label{g0pole} \end{figure}

\begin{figure} \caption{The one-pole contribution  to the spin
density, of the deuteron, $\langle \gamma_5\gamma_3\rangle
^{BS}_{pole}({\bf p})$, calculated with the BS amplitude.
Curves: the leading term in the Taylor expansion (dotted), first
two terms, up to $\propto p_2^2$ (dashed), first three terms, up
to $\propto p_4^2$ (solid).} \label{g53pole} \end{figure}


\begin{figure} \caption{Different ``versions'' of the charge
density  of the deuteron calculated with the BS amplitude.
Curves: exact, $\langle \gamma_0\rangle ^{BS}({\bf p})$ (dotted);
in the pole approximations, $\langle \gamma_0\rangle
^{BS}_{pole}({\bf p})$, (solid). Dashed curve presents effective
density for  deep inelastic scattering which is the density in
the pole approximation with account for the $\Theta$-function
cut-off at high momenta.  } \label{g0dis} \end{figure}

\begin{figure} \caption{The same as in Fig.~\ref{g0dis}, but for
the spin-density, $\langle \gamma_5\gamma_3\rangle ^{BS}({\bf
p})$.} \label{g53dis} \end{figure}


\begin{figure} \caption{The effective distribution for the
nucleon contribution in the deuteron structure function $F_2^D$,
$f^{N/D}(y)$. Curves: fully relativistic BS (solid);
nonrelativistic  Bonn (dotted), nonrelativistic, but using the BS
charge density  (dashed).} \label{fndp} \end{figure}


\begin{figure} \caption{The ratio of the deuteron and nucleon
structure functions, $F_2^D/F_2^N$, calculated in different
models. Curves correspond to three effective distributions from
Fig.~\ref{fndp}: fully relativistic BS (solid);  nonrelativistic
Bonn (dotted), nonrelativistic, but using the BS charge density
(dashed).} \label{rf2} \end{figure}

\begin{figure} \caption{The deuteron structure function
$F_2^D(x)$ at large $x$,  calculated in different models. Curves
correspond to three effective distributions from Fig.~\ref{fndp}:
fully relativistic BS (solid);  nonrelativistic  Bonn (dotted),
nonrelativistic, but using the BS charge density  (dashed).
Dash-dotted curve present the free nucleon structure function,
$F_2^N$.} \label{f2p} \end{figure}

\begin{figure} \caption{The effective distribution functions for
the deuteron structure functions, $b^D_{1,2}$, $\Delta
f^{N/D}(y)$, calculated  in different models.  Curves: fully
relativistic BS (solid);  nonrelativistic  Bonn (dotted). The
dashed curve presents $y \Delta f^{N/D}(y)$.} \label{dfndp}
\end{figure}

\begin{figure} \caption{The deuteron structure functions
$b^D_{1}$ (a) and  $b^D_{2}$ (b), calculated  in different
models. Curves: fully relativistic BS (solid);  nonrelativistic
Bonn (dashed). The dotted curve presents $b^D_{1,2}$ calculated
with the  ``soft'' nucleon distribution, the Bonn distribution
but cut-off at $|{\bf p }| > 0.7$~GeV .} \label{b2p} \end{figure}

\begin{figure} \caption{The ratio of the deuteron and nucleon
spin-dependent structure functions, $g_1^D/g_1^N$, calculated  in
different models. Curves correspond to three effective
distributions: fully relativistic BS (solid);  nonrelativistic
Bonn (dotted), nonrelativistic, but using the BS spin density
(dashed).} \label{g1} \end{figure}

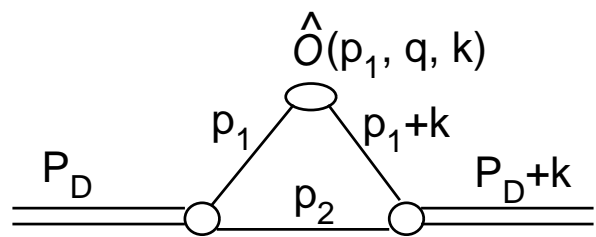
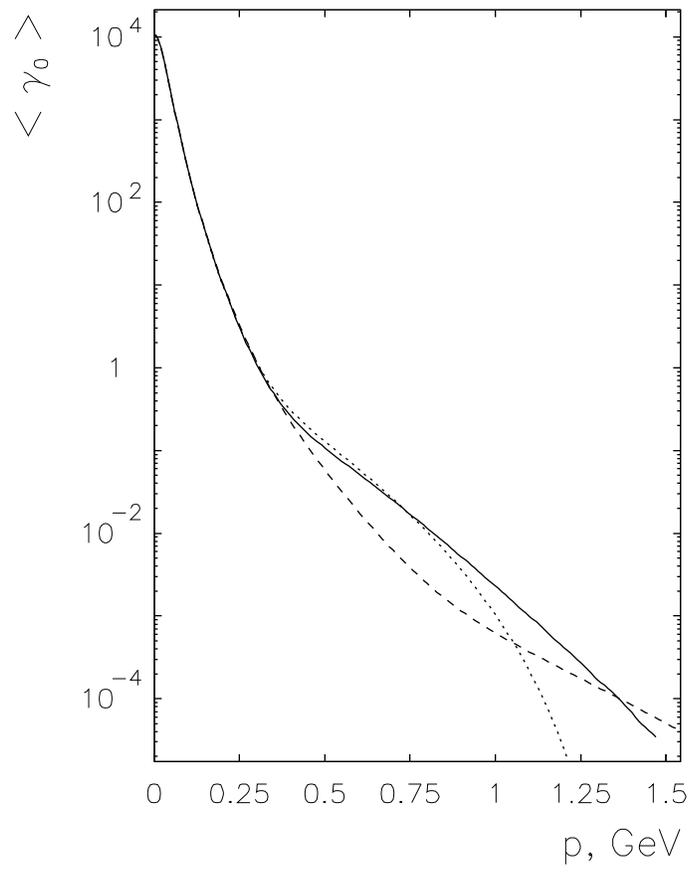
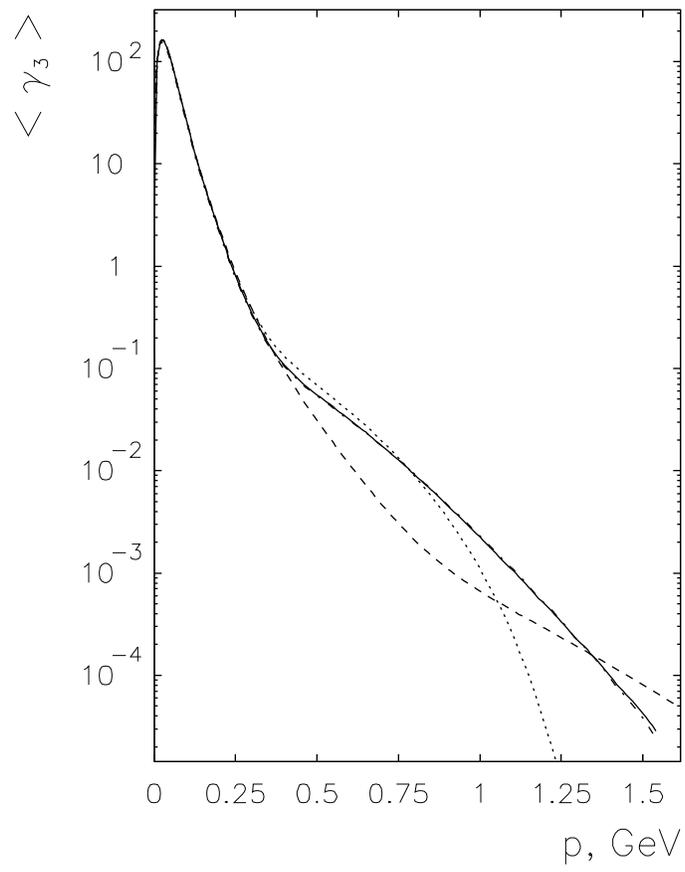
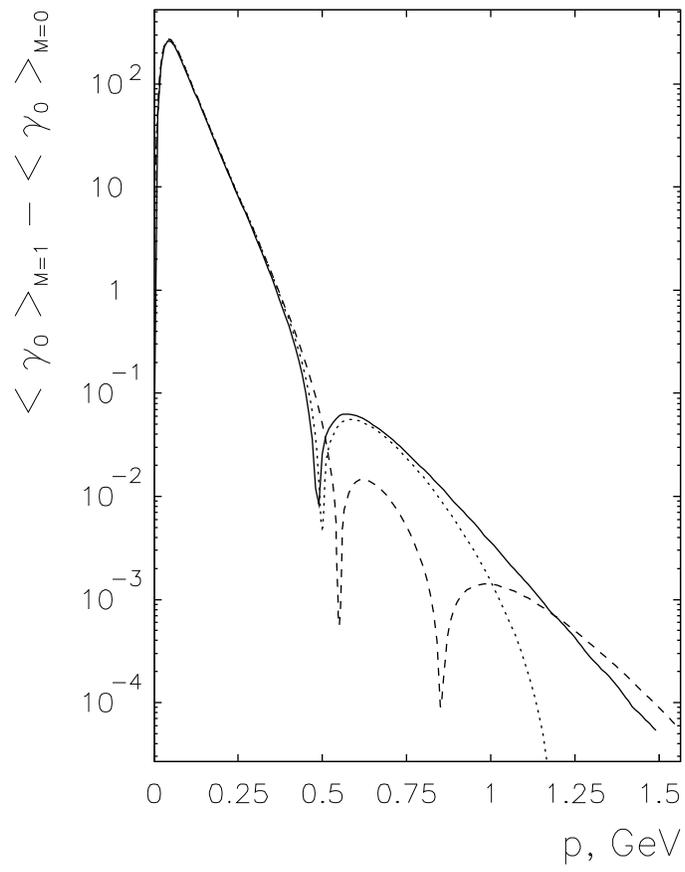
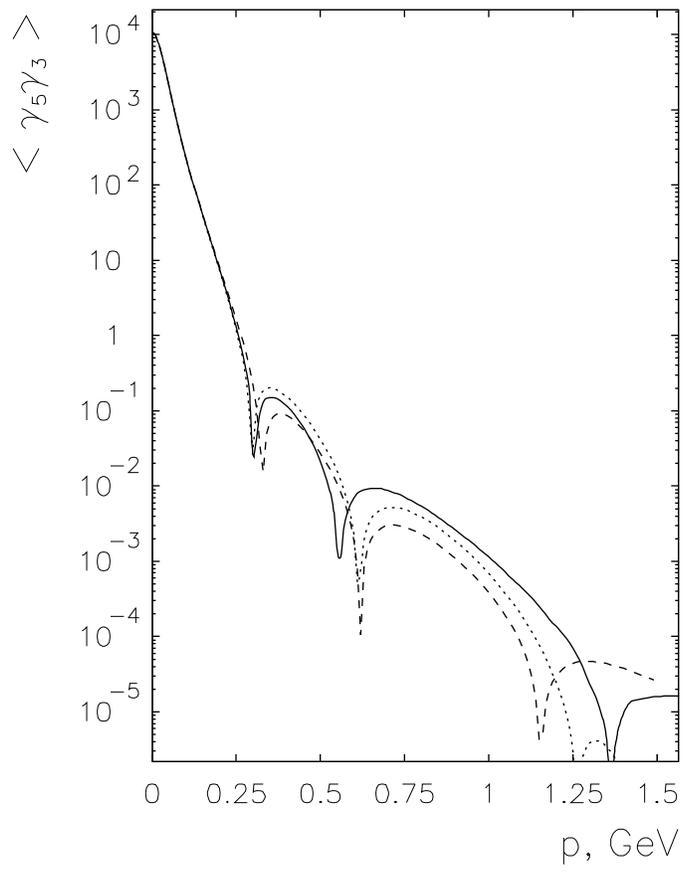
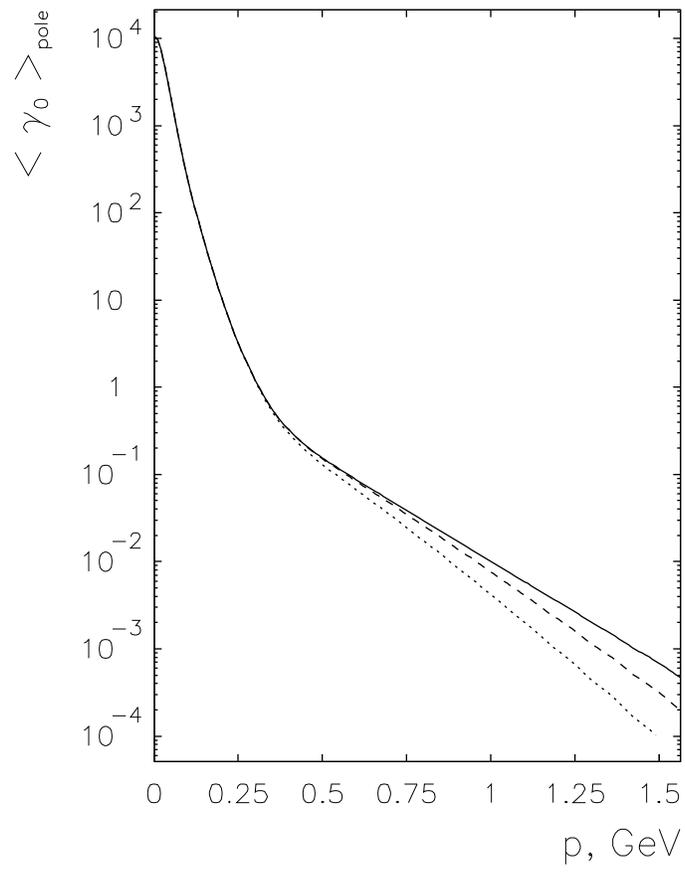
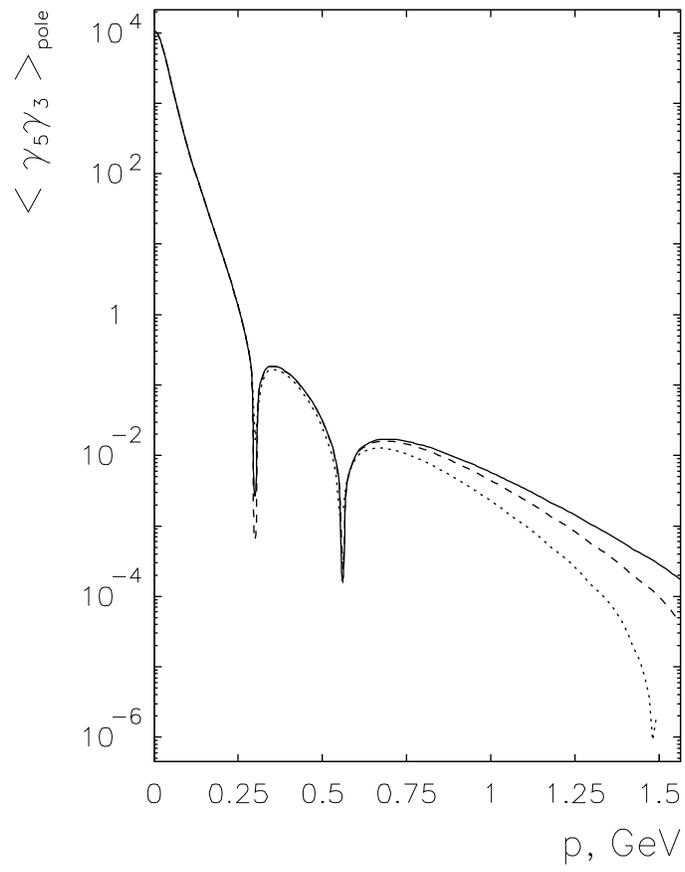
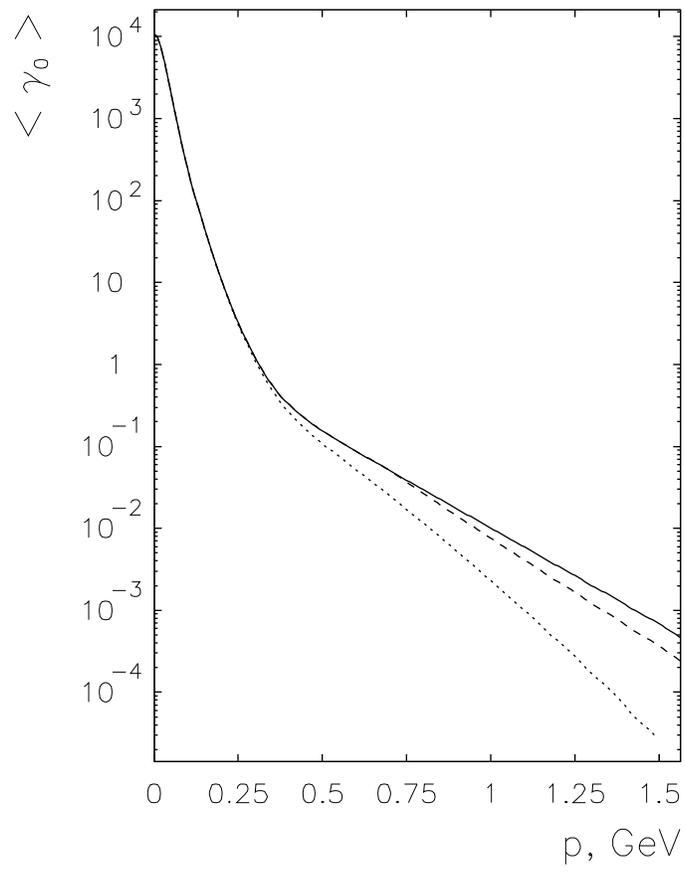
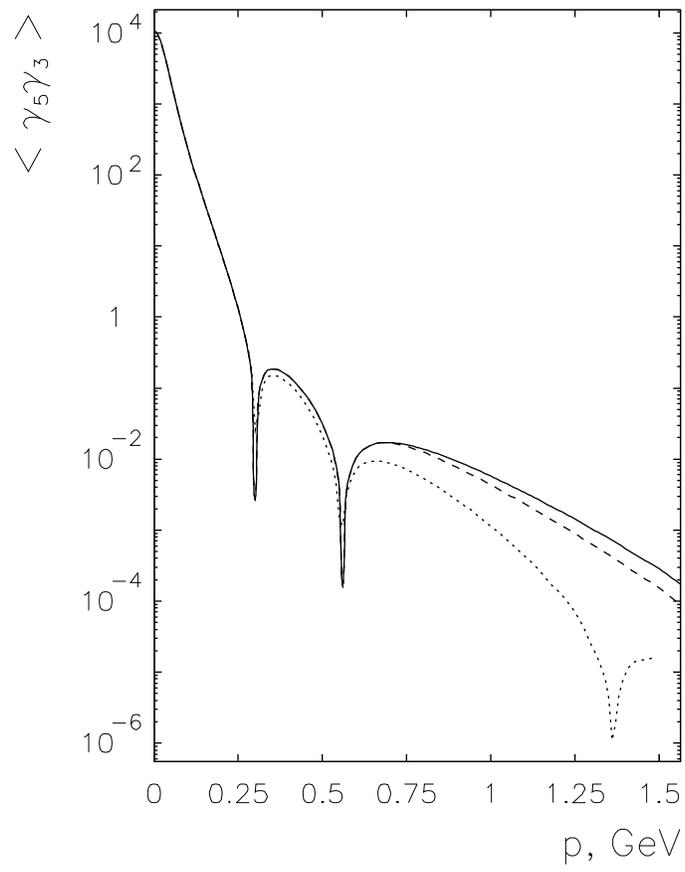
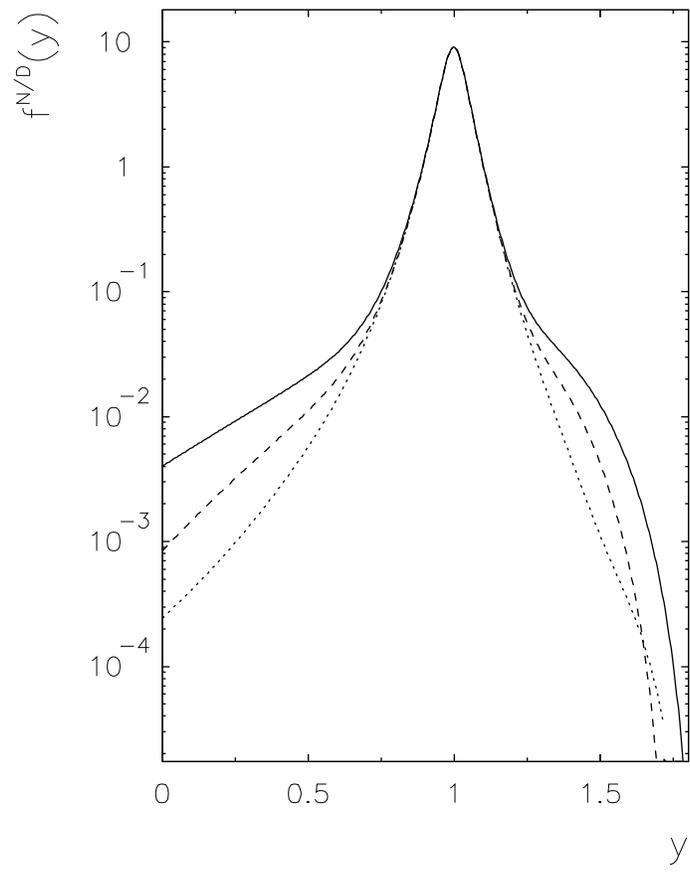
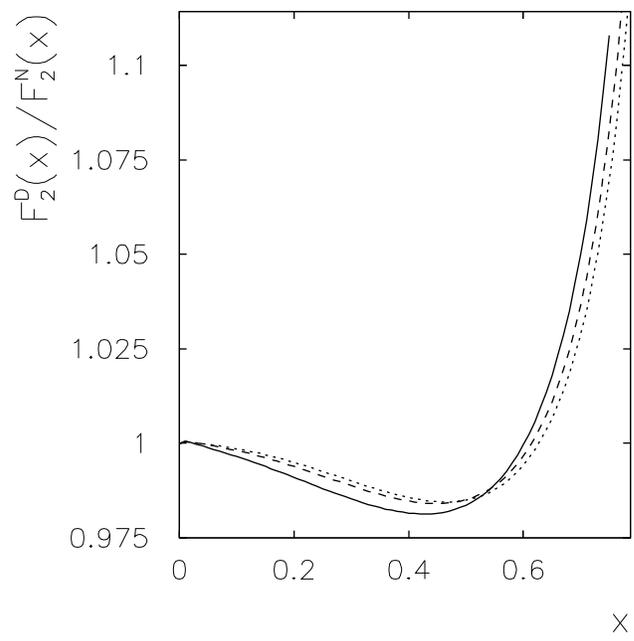
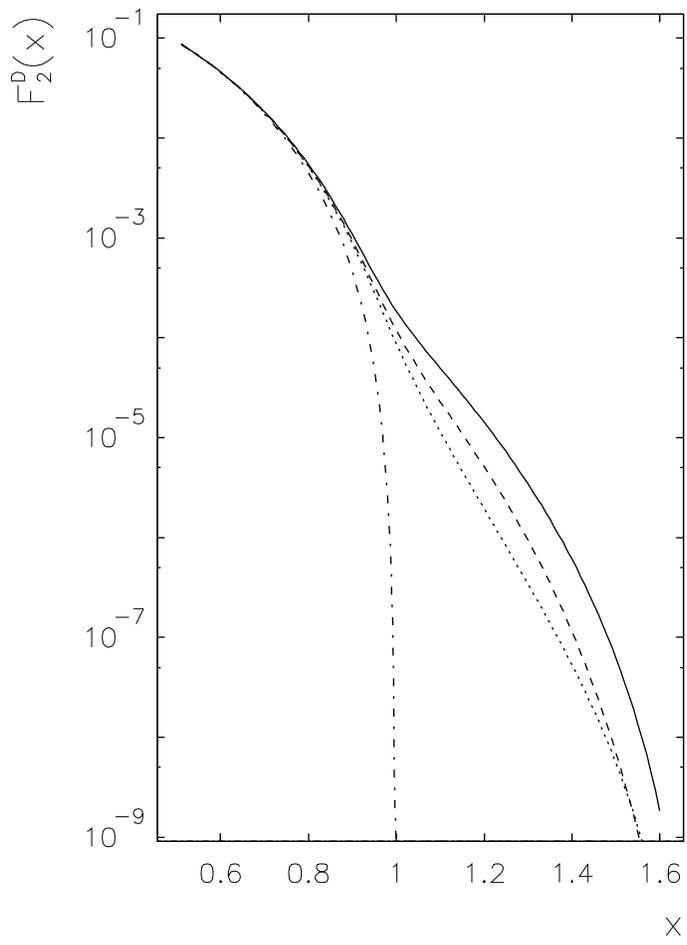
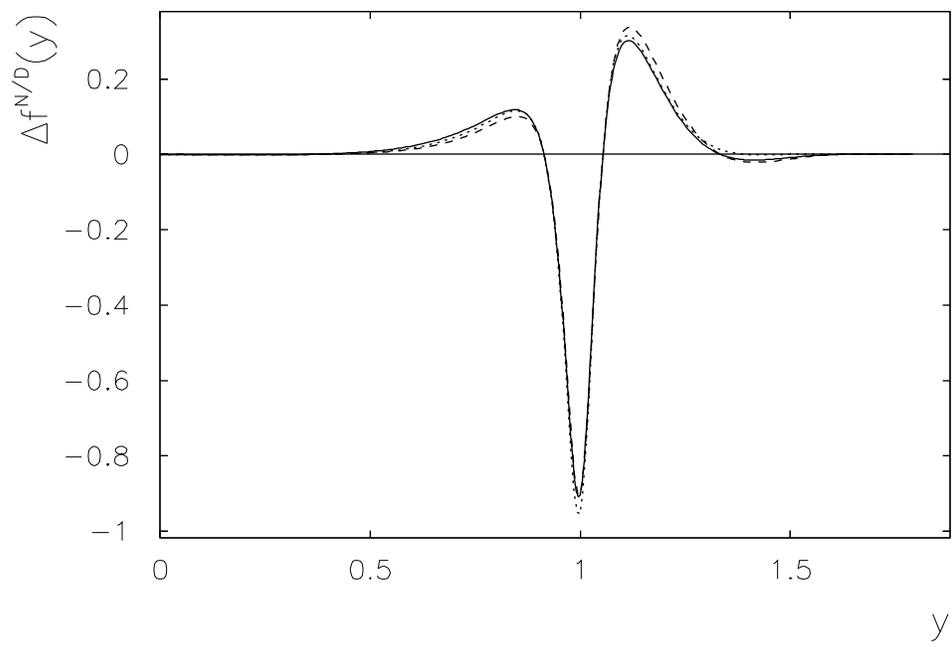
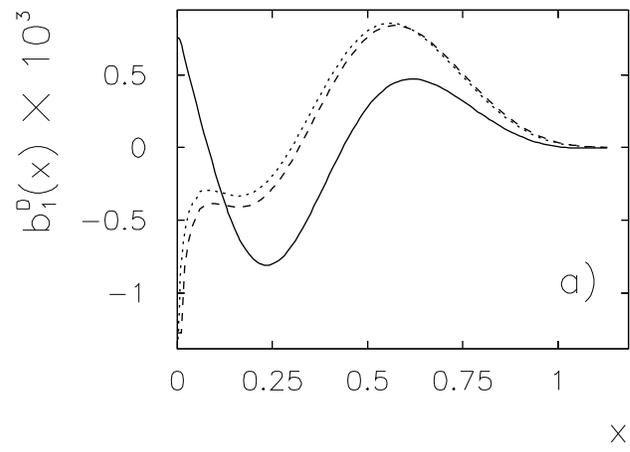
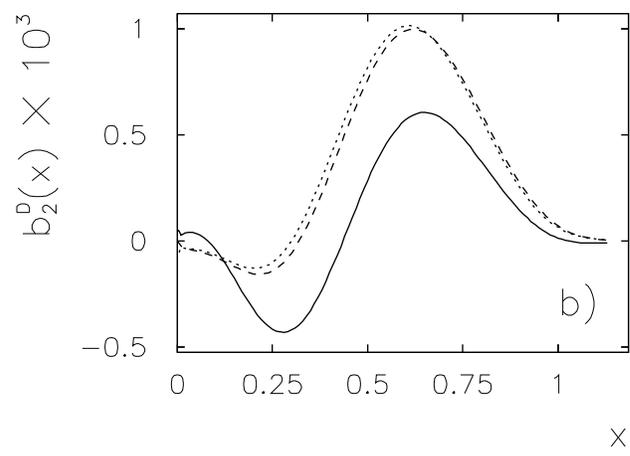
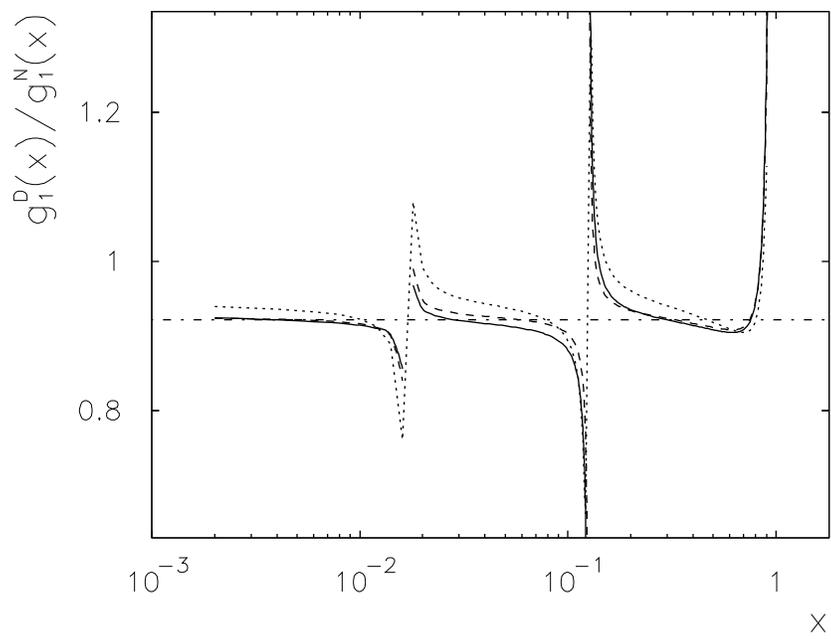
\end{document}